\documentclass[trackchanges]{aastex701}

\begin{document}

\title{Mid-Infrared Variability as a Probe of Hidden AGN Activity: A Comparison of Seyfert 1 and Seyfert 2 AGNs in the BAT 58-Month Sample}

\author[0009-0002-4279-9925]{Natalie Kovacevic}
\affiliation{Homer L.\ Dodge Department of Physics and Astronomy,
University of Oklahoma, Norman, OK 73019, USA}
\email{natalie.kovacevic@ou.edu}

\author[]{Lorelei Starling}
\affil{Homer L.\ Dodge Department of Physics and Astronomy,
University of Oklahoma, Norman, OK 73019, USA}
\email{Lorelei.P.Starling-1@ou.edu}

\author[0000-0001-9203-2808]{Xinyu Dai}
\affil{Homer L.\ Dodge Department of Physics and Astronomy,
University of Oklahoma, Norman, OK 73019, USA}
\email{xdai@ou.edu}

\author{Emilia Jaervelae}
\affiliation{Department of Physics and Astronomy, Texas Tech University, Box 41051, Lubbock, TX, 79409-1051, USA}
\email{}

\begin{abstract}

In accordance with the AGN Unification model, the optical variability of Seyfert 2s is expected to be suppressed because the central engine is obscured by a dusty torus. The absorbed UV/optical radiation is reprocessed and re-emitted in longer wavelengths including the mid-infrared (MIR), making MIR variability a promising probe of obscured AGNs. We investigate the optical ASAS-SN and MIR NEOWISE variability of Seyfert 1 and Seyfert 2 AGNs from the \textit{Swift} 58-month BAT catalog, comparing both populations against non-variable galaxies. One of the first systematic studies of MIR AGN variability, we detected significant variability for Sy1s and the majority of Sy2s. 65\% of Sy2s exhibit variability as significant as Sy1s in the MIR, especially in the W2 band, indicating torus dominated MIR variability.  
Additionally, 22\% of Sy2s show higher variability compared to non-variable sources in the W2 band with p-value of $7.1\times 10^{-15}$. Demonstrating MIR variability as an important method to probe the variability of torus and select obscured AGN.
Finally, we perform dust reverberation mapping on the Sy1 sample between the optical and W1 or W2 light curves, with measured time delays between 22 -- 786 days (0.02--0.61 pc) for
the hot dust in the inner regions of the torus.
Our measured torus radius-luminosity slope, $R \propto L^{0.42\pm0.05}$, agrees with literature results. 
Combining the W1 and W2 band lags with archival NIR measurements, we measured the temperature profile of the torus with $T \propto R^{-\alpha}$, where $\alpha$ values range from 0.3 -- 1.1.

\end{abstract}


\section{Introduction} 

Active galactic nuclei (AGNs) are luminous sources in our universe found in the center of most massive galaxies powered by the accretion of material onto a central supermassive black hole (SMBH), visible throughout the electromagnetic spectrum. Seyfert galaxies (Sy) a type of AGN often observed in the local universe, display typical bolometric luminosities between $10^{43}$--$10^{45}$ ergs s$^{-1}$, making them less luminous and nearby compared to QSOs (Quasi-Stellar Objects). The unification model was introduced to classify AGNs based on their viewing angle to the observer (\citealt{Antonucci_1993ARA&A..31..473A,Urry_1995PASP..107..803U}). This model consists of the central engine, the broad line region (BLR) surrounded by a large scale dust torus, and an extended narrow line region (NLR). Sy1 galaxies have a direct view to the central engine, visibly showing the entire optical emitting region and resulting in the presence of broad and narrow optical emission lines in their spectra. The BLR of Sy2s is obscured by the dust torus allowing only the narrow lines to be present in their spectra.

A key feature of AGNs is their stochastic variability, that is observed at all wavelengths and timescales. The physical mechanism responsible for optical variations is unclear, with possible origins as X-ray reprocessing, accretion disk instabilities, star collisions, and supernova
explosions in host galaxies or accretion disk. While the optical and X-ray variability probe the accretion disk and corona property changes, the dominant physical process responsible for infrared variability is the thermal reprocessing of UV/optical photons by the dust torus (\citealt{Nenkova_2008,mandal_10.1093/mnras/sty200}). Galaxies lacking an active SMBH intrinsically exhibit no infrared variability as their continuum is dominated by the non-variable ensemble of stars and dust of the host galaxy. The orientation of Sy1s, having a clear line of sight of the central engine and the dust torus at all radii, making it ideal for optical and infrared variability studies. On both the long and short time scale these studies reveal that Sy1s exhibit large variations in the optical and near infrared (NIR) band with the most prominent variations occurring at the long timescale (\citealt{deRuiter_1986A&AS...63...59D,lopez2023type}). Sy2s in contrast have an obscured view of the central engine, making variability studies difficult. For the decade long timescale, \citet{kovacevic_2026ApJ..1006..130K} observed Sy2s to exhibit significant optical variability beyond the constant flux established by galaxies. This raises the question on whether we see larger variability in the long baseline infrared light curves from the reprocessed optical emission.

The mid-infrared (MIR), spanning approximately 2.2--25~$\mu$m, probes dust in the inner to cooler, more extended regions of the dusty torus. Despite its importance for understanding the structure and variability of AGN dust emission, relatively few studies have investigated AGN variability across this wavelength range. Using a sample of both radio-loud and radio-quiet quasars, \citet{Neugebauer_1999} found that the majority of their sample exhibited measurable variability from the NIR through the MIR at 10~$\mu$m. More recently, \citet{Izviekova_2026arXiv260506517I} examined a small sample of narrow-line Sy1 galaxies and found that the variability amplitudes of their ZTF $g$-band light curves were larger than the smoothed variability amplitudes observed in the NEOWISE W1 band. Similarly, \citet{lyu_2019ApJ...886...33L} investigated the variability of low-redshift quasars from the optical through the MIR and found that the variability amplitude was largest at optical wavelengths and decreased toward the MIR.
From a comparison study of the MIR variability of broad and narrow line Sy1s \citep{Hu_article}, they were observed to differ at the longest time scale. Further analysis of their ensemble structure functions revealed for the broad line Sy1s a positive correlation with black hole mass and for both the broad and narrow line Sy1s a negative correlation with eddington ratio.

Among the various diagnostics of time-domain studies, reverberation mapping is a technique used to probe the inner structure of AGNs by measuring the time delay between variability in the driving continuum and the response of surrounding emitting material \citep{blandford_1982ApJ...255..419B}. In dust reverberation mapping, variations in the optical/UV emission from the accretion disk heat the surrounding dust, which subsequently re-emits this energy at infrared wavelengths after a measurable time delay. This technique was first applied to the Sy1 galaxy Fairall 9, where variations in the optical/UV continuum were followed by a near-infrared response with a time lag of approximately 400 days \citep{clavel_1989ApJ...337..236C}. Subsequent studies have used dust reverberation mapping to measure the radius of the hottest dust probed at near-infrared wavelengths and to investigate the relationship between dust radius and AGN luminosity \citep{Minezaki_2019,pozo_2025A&A...700L...8P}. As the dust temperature decreases with increasing distance from the central source, observations at longer infrared wavelengths can probe cooler dust at larger radii. Measuring reverberation lags across multiple infrared wavelengths can therefore be used to investigate the temperature profile and radial structure of the dusty torus.

In this study we analyze the variability of Sy1 and Sy2 galaxies in the optical and MIR wavelengths. Performing one of the first comprehensive MIR variability studies on AGNs, our results show that the variable re-emitted optical light of Sy2s is detected and at larger radii are largely indistinguishable from Sy1 variability. This was concluded from comparing the normalized excess variance of the Sy1 and Sy2 sample against a sample of constant non-active galaxies. From our results we also perform a dust reverberation mapping study on our Sy1 sample and calculated the size of the inner regions of the torus, where are measurements are consistent with other reverberation mapping studies and constrained the temperature profile of the torus for a sample of Sy1s.
Throughout the paper we assumed cosmological parameters of H$_0$ = 72 km s$^{-1}$Mpc$^{-1}$, $\Omega_m$ = 0.3, and $\Omega_\Lambda$ = 0.7.
\section{Sample Selection and Data} \label{}

\subsection{\textit{Swift-BAT 58 Month Survey and 2MASS Extended Source Catalog}}

Our sample expands on the work done by \cite{kovacevic_2025ApJ...985..177K} studying type 1 and type 2 AGNs in the \textit{Swift} Bat 9 month survey \citep{Tueller_2008ApJ...681..113T} to include the fainter \textit{Swift} Bat 58 month survey \citep{Baumgartner_2010HEAD...11.1305B}. This is the first 58 months of the hard X-ray all sky survey on the Neil Gehrels Swift Observatory \citep{Gehrels_2004ApJ...611.1005G} using the Burst Alert Telescope \citep{Barthelmy_2005SSRv..120..143B}, building upon the previous brighter 22 month and 9 month surveys. We also excluded any Seyferts that are classified as radio loud as this can add a level of variability from jets. Our final sample includes 165 Sy1s and 161 Sy2s, where Sy1.2 and Sy1.5 are grouped as unobscured Sy1s and Sy1.8s and Sy1.9s are grouped as obscured Sy2s based on the relative strength of the broad H$\beta$ emission line, with redshifts ranging from 0.001--1.554. 

The sample of galaxies used in this study as the non-variable comparison sample was derived from the Two Micron All Sky Survey (2MASS) \citep{skrutskie_2006AJ....131.1163S} extended source catalog \citep{Jarrett_2000AJ....119.2498J}. Following the same initial selection as \cite{Yuk_2026arXiv260323752Y} to include bright galaxies (K $<$ 11.34) away from the galactic plane $|b|$ $>$ 15$^{\circ}$. Another round of filtering was performed after extraction of the ASAS-SN light curves, including a V-band magnitude cut (11.5 $<$ V $<$ 16.5) and a cut based on the flux uncertainties of the ASAS-SN light curves, resulting in 40 galaxies in our final sample.

\subsection{NEOWISE Mid-Infrared Light curves and ASAS-SN Optical Light curves}

The data used in this study comes from the Near-Earth Object Wide-field Infrared Survey Explorer (NEOWISE; \citep{mainzer_2014ApJ...792...30M}) and The All-Sky Survey for SuperNovae (ASAS-SN \citealt{Shappee_2014,hart_2023arXiv230403791H}). The two surveys almost cover contemporaneously over a decade allowing for an ideal variability comparison of the MIR and optical light curves.

The Wide-field Infrared Survey Explorer \citep[WISE,][]{wright...2010AJ....140.1868W} was a space telescope that surveyed the entire sky in 4 mid-infrared wavelengths, 3.4, 4.6, 12, and 22 $\mu$m (referred to as W1, W2, W3, and W4, respectively) from January 7 to August 6, 2010. 
After the cryogen was exhausted, the survey was continued as NEOWISE, continuing observations in the W1 and W2 wavelengths, first to the inner solar system until its hibernation in 2011. 
NEOWISE was brought out of hibernation in 2013 and decommissioned in August 2024. From 2013 to 2024 NEOWISE has performed repeated all-sky surveys, with each all-sky survey lasting approximately six months.
In each all-sky survey, NEOWISE achieved at least 12 independent single exposures on each point in the sky within 1.5 days.
The magnitude limit of these single exposures are $15.0$ mag in W1 and $13.0$ mag in W2 with a signal-to-noise of 10. 
There exist dependable detections that are fainter than these limits, but the rate of occurrence of unreliable detections will be greater. Sources brighter than 8.0 mag in W1 and 7.0 mag saturate the detectors, which is flagged in the data. 

Light curves were extracted by querying the NEOWISE-R Single Exposure (L1b) Source Table through the IRSA archive using the Python package Astroquery. Using coordinates procured from the Swift BAT 58-Month Hard X-ray Survey, we performed a cone search with a radius of 1.5 arcseconds around the locations of each source. We discarded sources that did not return NEOWISE data and furthered filtered the returned data based on quality flags, including contamination and confusion, frame quality score, frame image quality score, angular separation on the sky from the South Atlantic Anomaly, and moon masking. 

Since the single exposure flux measurements exhibit systematic uncertainties, we binned the single exposure observations such that each all-sky visit yields one measurement. 
We binned the single exposure measurements in the flux unit, excluded outliers, calculate the mean and standard deviation of the remaining measurements, and converted back to magnitudes.
Epochs with less than 5 single exposures are excluded to ensure reliable estimates of the average fluxes and uncertainties.

\begin{figure}[ht]
\centering
\includegraphics[width=0.48\textwidth,trim={0cm 0cm 0cm 0cm}]{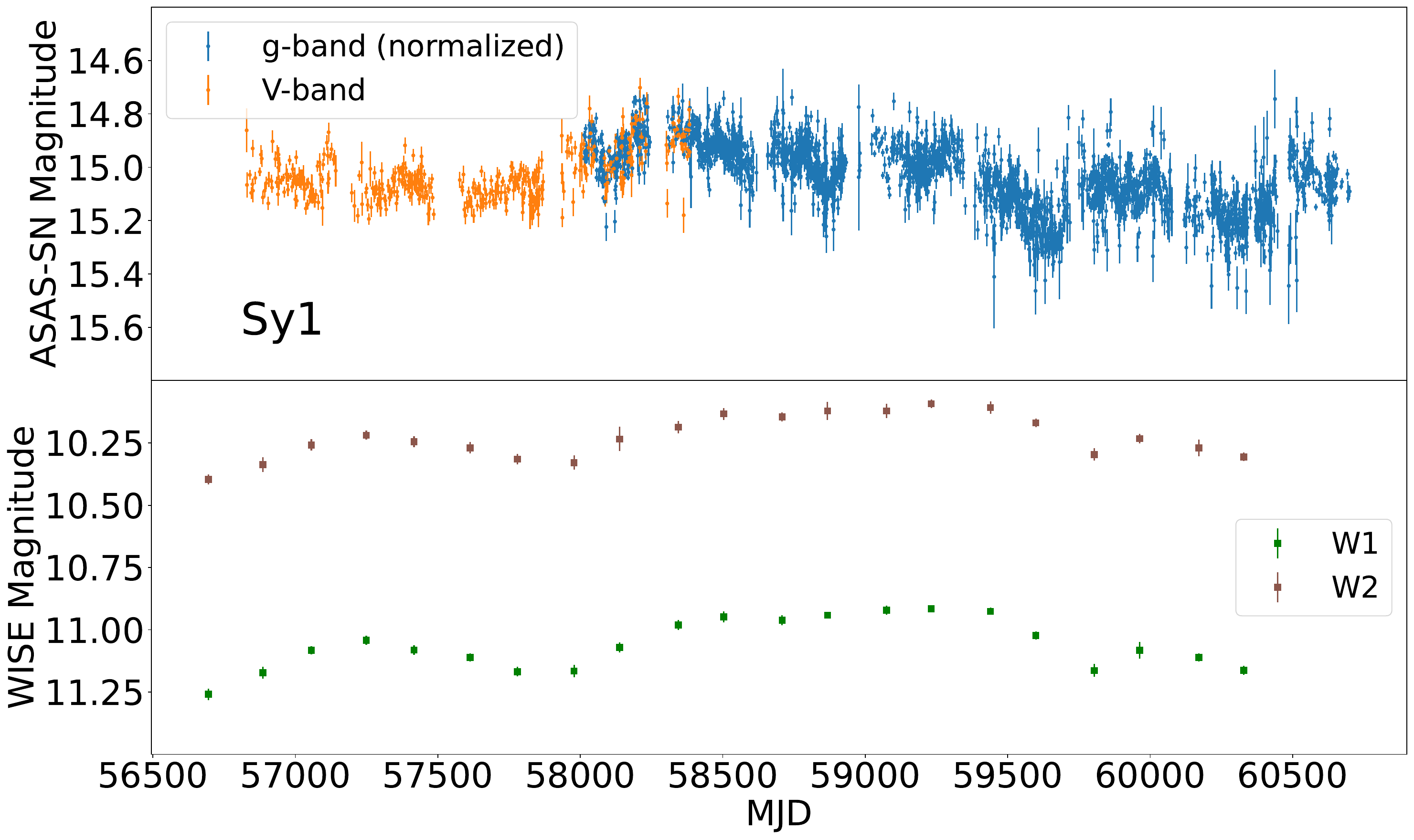}
\includegraphics[width=0.48\textwidth,trim={0cm 0cm 0cm 0cm}]{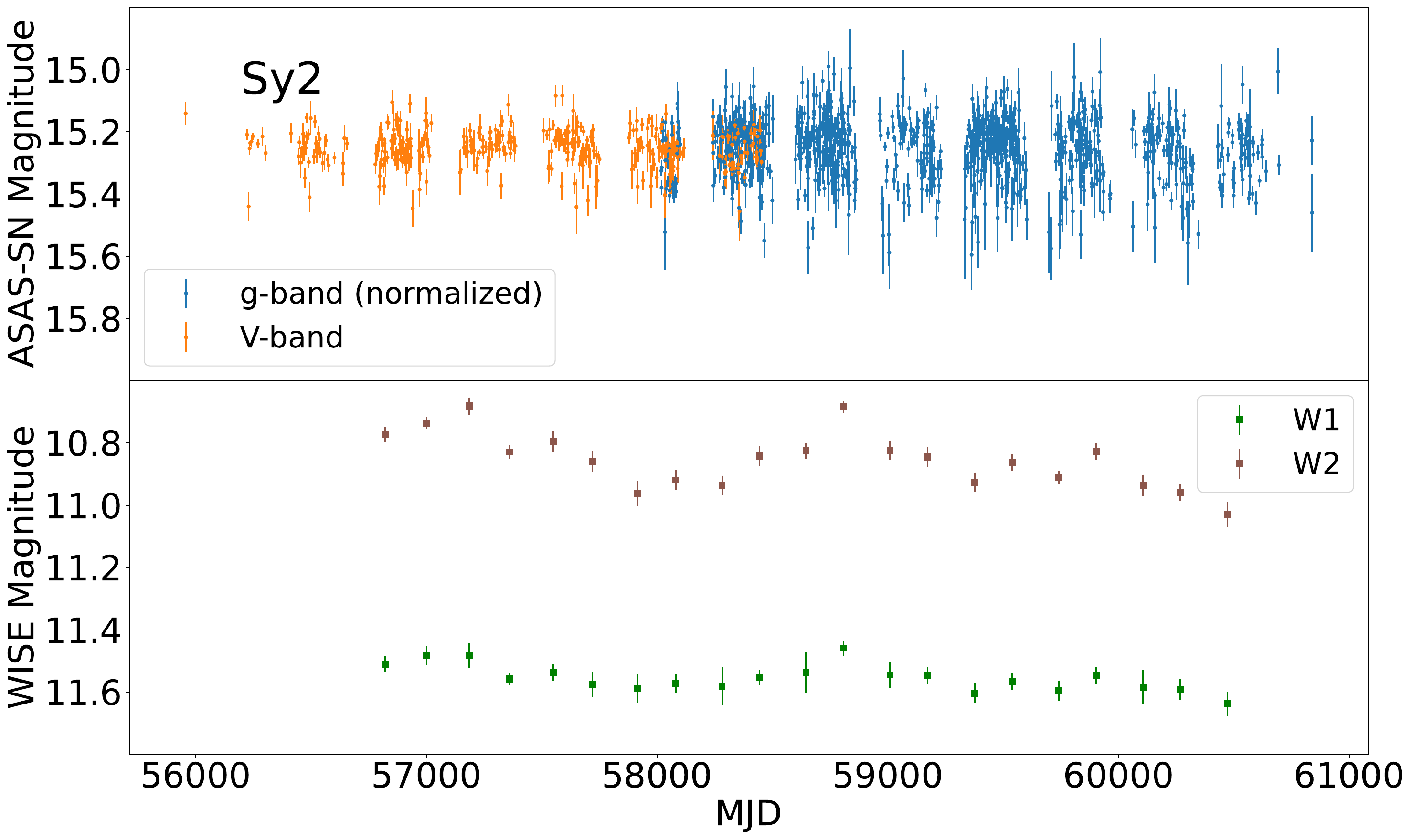}
\includegraphics[width=0.48\textwidth,trim={0cm 0cm 0cm 0cm}]{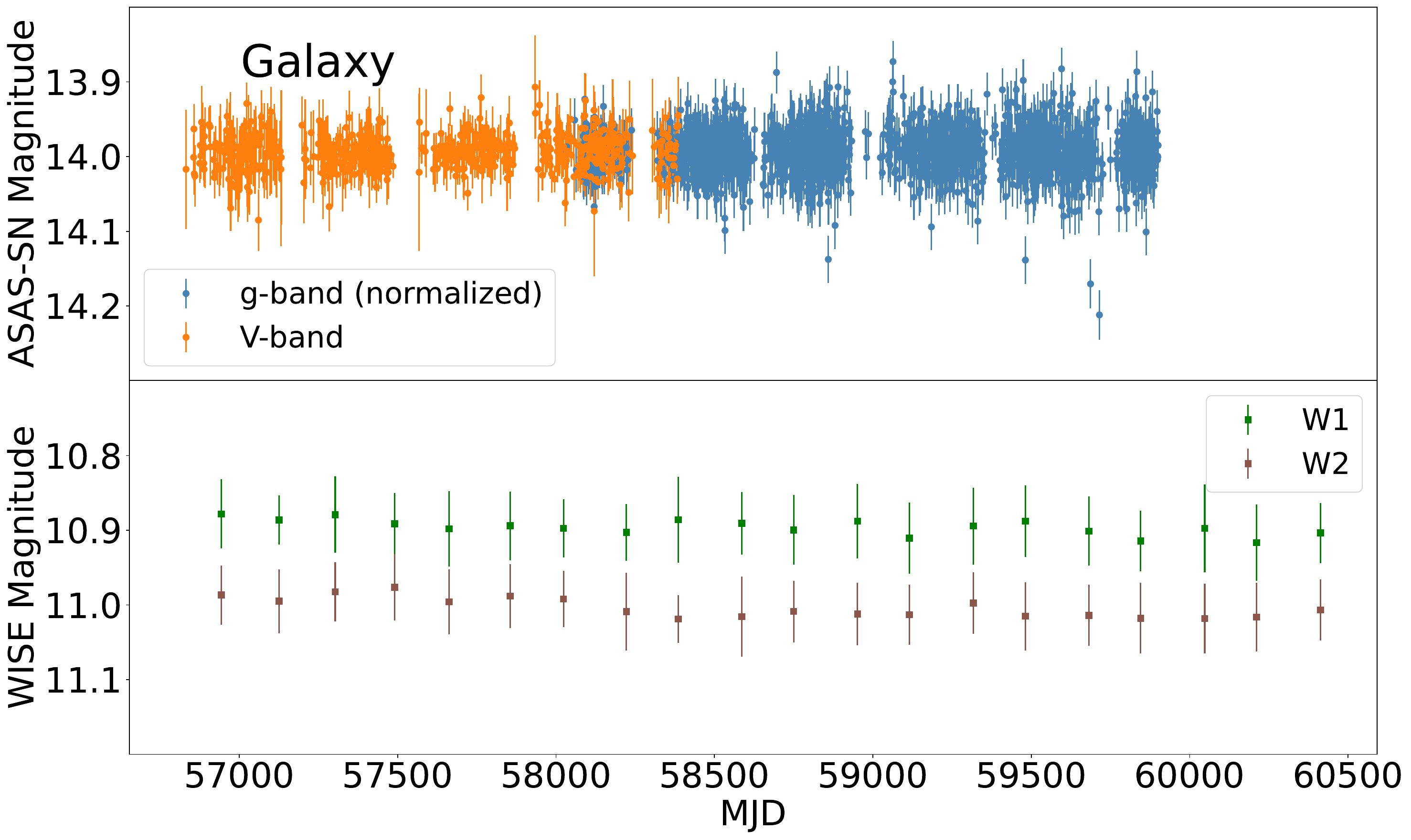}
\caption{Top left: Example ASAS-SN and NEOWISE of the Sy1 2MASXJ04372814-4711298 displaying clear stochastic variability in both optical and Mid-IR. Top right: Example optical and Mid-IR light curve of the Sy2 UGC12237 exhibiting noticeable variability in the Mid-IR. Bottom: Example ASAS-SN and NEOWISE of the galaxy ESO 33-4 establishing the constant flux baseline.}\label{lightcurves}
\end{figure}

In contrast, ASAS-SN (\citealt{Shappee_2014,kochanek_2017PASP..129j4502K}) is a ground-based optical survey composed of multiple telescopes that has monitored the visible night sky continuously for more than a decade from April 2013 to present day. Its sensitivity, reaching a limiting magnitude of approximately 18.5 in the $g$ band, provides sufficiently long and well-sampled baselines for constructing light curves for our sample \citep{kovacevic_2026ApJ..1006..130K}. We obtained the ASAS-SN photometry through image subtraction followed by aperture photometry, with the resulting measurements calibrated relative to an ensemble of comparison stars within each field (\citealt{Jayashinghe_10.1093/mnras/stz444,jayasinghe_10.1093/mnras/sty838}). Utilizing this extraction method resolves the nucleus variability.
The dataset spans several generations of the ASAS-SN instrumentation. Observations initially came from two mounts operating with the $V$-band filter. These were subsequently expanded to five mounts, incorporating both $V$- and $g$-band filters, before the original mounts were ultimately replaced with $g$-band systems. During the transition between the two filters, approximately 400 days of overlapping observations were available. We used the mean flux offset measured over this overlap period to place the $V$- and $g$-band measurements on a common flux scale and produce a normalized combined light curve. Examples of both the Mid-IR and optical Sy1, Sy2, and galaxy light curves are shown in the left, right, and bottom panels of Figure~\ref{lightcurves}, respectively.

\section{Analysis: Normalized Excess Variance}

The normalized excess variance is a commonly used metric in AGN physics to quantify the variability of AGN light curves (\citealt{vaughan_2003MNRAS.345.1271V,schleicher_2019Galax...7...62S}). This metric removes the contributions of uncertainties in the flux measurement $\sigma_{\textrm{err}}$ from the variance $S^2$, which is then normalized by the mean of the light curve $\langle{x}\rangle$.

\begin{equation}
\sigma_{\textrm{\tiny{NXS}}}^2 = \frac{S^2 - \langle{\sigma_{\textrm{err}}^2 }\rangle}{\langle{x}\rangle^2} \label{eq1}
\end{equation}

The error on the normalized excess variance is determined by

\begin{equation}
\Delta{\sigma_{\textrm{\tiny{NXS}}}^2} = \sqrt{\Biggl( \sqrt{\frac{{2}}{N}}\frac{\langle{\sigma_{\textrm{err}}^2 }\rangle}{\langle{x}^2\rangle} \Biggr)^2 + \Biggl({\sqrt{\frac{\langle{\sigma_{\textrm{err}}^2 }\rangle}{N}}\frac{2F_{\textrm{var}}}{\overline{x}}}\Biggr)^2} \label{eq2}
\end{equation}

where N is the number of data points in the light curve and $F_{\textrm{var}}$ is the fractional variability. 

Any source that returned a negative normalized excess variance value was treated as having zero variability.

\begin{equation}\label{equation3}
F_{\textrm{var}} =\sqrt{\sigma_{\textrm{\tiny{NXS}}}^2}
\end{equation}

The fractional variability is a preferred metric for the comparison between different energy bands, as it is a fractional root-mean square expressing the variability in terms of a fraction. 

\section{Results}

\subsection{Variability of Sy2s in Mid-Infrared}
\begin{figure}[ht]
\centering
\includegraphics[width=0.405\textwidth,trim={0cm 0cm 0cm 0cm}]{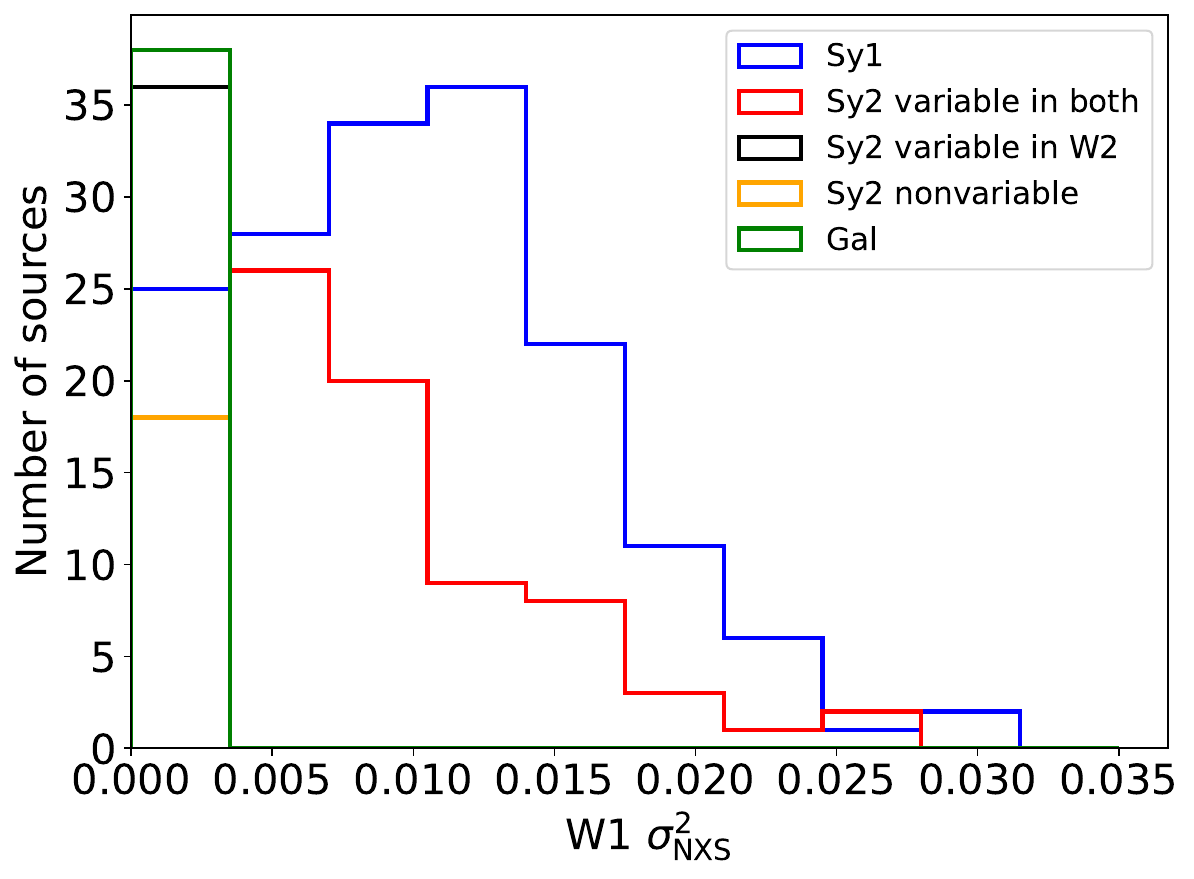}
\includegraphics[width=0.4\textwidth, trim={0cm 0cm 0cm 0cm}]{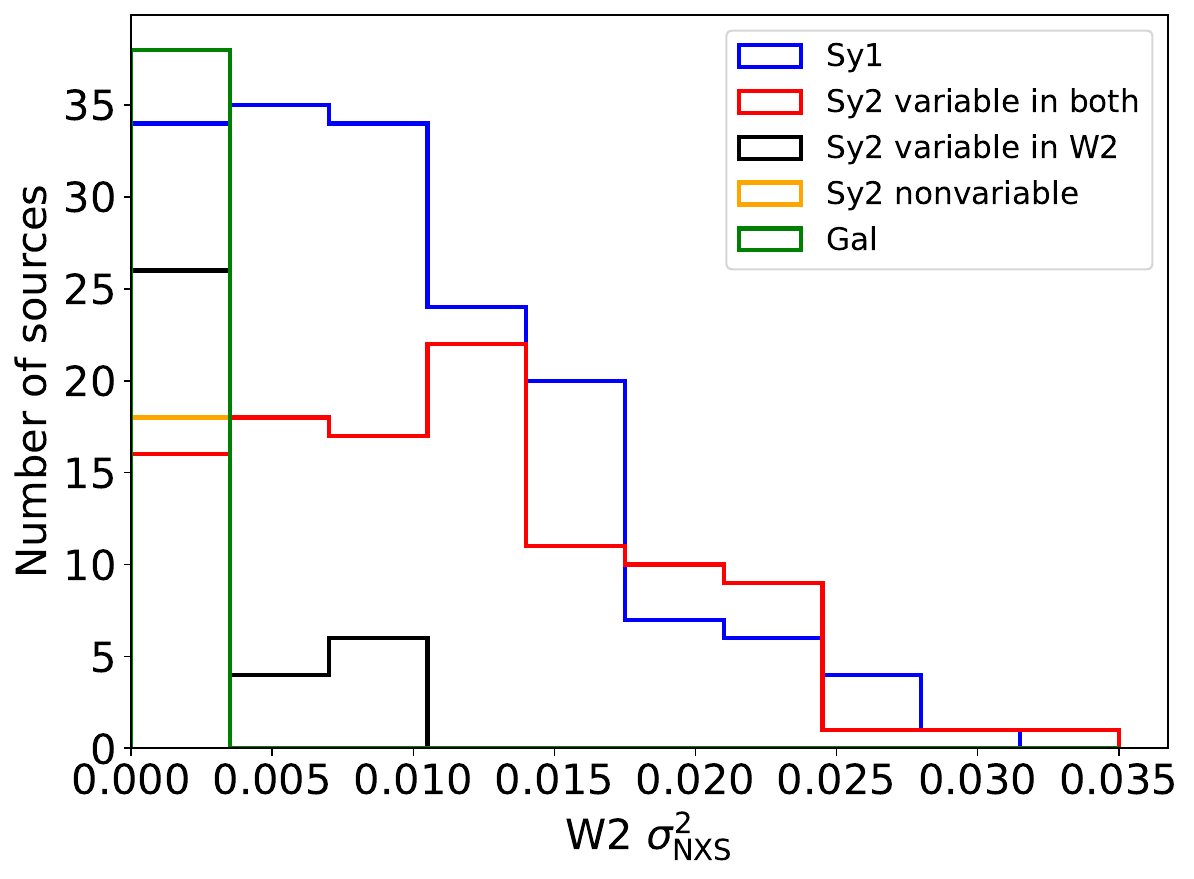}
\caption{Distributions of the normalized excess variance for the different population in W1 and W2 band. The mean $\sigma_{\textrm{\tiny{NXS}}}^2$ values for Sy1s, Sy2s (variable in both), Sy2s (variable in W2) and Sy2s (nonvariable) are 0.01, 0.007, $7.9\times 10^{-5}$and 0.0 for the W1 band (left Panel). The mean $\sigma_{\textrm{\tiny{NXS}}}^2$ values for the above listed types in the W2 band (right panel) are 0.01, 0.01, 0.003, and $5.2\times 10^{-5}$. This suggests similar variability for the Sy1 and Sy2 variable in both samples in the W2 band.}
\label{histogram_wise}
\end{figure}

To compare the variability of the three populations (Sy1s, Sy2s, and Galaxies), in Figure~\ref{histogram_wise} we show the histograms of the normalized excess variance of the populations in both W1 and W2 bands, where a substantial fraction of Sy2s share similar variability amplitudes as Sy1s. We performed a  Kolmogorov-Smirnov (KS) test on the Sy1 and Sy2 sample yields p-value: $7.9\times 10^{-14}$ and $1.6\times 10^{-12}$ for W1 and W2 respectively. We further separate the Sy2 population based on the level of variability,  ($\sigma^2_{\textrm{\tiny{NXS}}} - \Delta{\sigma_{\textrm{\tiny{NXS}}}^2} > 0$), in three categories:
variable in both W1 and W2 bands, variable in W2 band only, and variable in neither band with 106, 36, and 19 Sy2s in these categories.
There are no Sy2s that are detected as variable in W1 only in our sample.
 The mean $\sigma^2_{\textrm{\tiny{NXS}}}$ value for Sy1s and Sy2s in each category in the W1 are Sy1s $\sigma_{\textrm{\tiny{NXS}}}^2 = 0.01$, Sy2s variable in both bands $\sigma_{\textrm{\tiny{NXS}}}^2 = 0.007$, Sy2s variable in W2 only $\sigma_{\textrm{\tiny{NXS}}}^2 = 7.9\times 10^{-5}$, and Sy2s nonvariable $\sigma_{\textrm{\tiny{NXS}}}^2 = 0.0$. In the W2 band the mean values are 0.01, 0.01,  0.003, and 5.2$\times 10^{-5}$ for the above four categories respectively.
The mean $\sigma^2_{\textrm{\tiny{NXS}}}$ values of variable Sy2s in both bands are approching those of Sy1s.
We performed more sensitive KS-tests on the normalized excess variance distributions, where the null hypothesis threshold of p $<$ 0.01 was used to further distinguish different populations \citep{Wasserstein02042016}. 
Between the Sy1 and Sy2 variable in both samples, the KS-test yield p-values of 3.3$\times 10^{-5}$ and 0.04, respectively, in the W1 and W2 bands, suggesting that Sy1s are more variable than Sy2s (variable in both bands) in W1. In W2 the variability amplitudes between Sy1s and variable Sy2s are not significantly distinguished.
Combining the KS result with the mean values in W2 the Sy2s (variable in both bands) sample exhibits levels of variability similar to that of Sy1s. When we compare the nonvariable Sy2s to our galaxy sample the KS test yields p-values of 1.0 in the W1 band and 0.1 in W2, signifying that nonvariable Sy2s are indistinguishable from those of constant galaxies. Previous optical studies have shown that Sy1s are more variable than Sy2s in the optical band (\citealt{lopez2023type,kovacevic_2025ApJ...985..177K}), and that Sy2 variability is even lower than that of star-forming and composite galaxies in some samples \citep{yip2009probing}. However, in a recent survey \cite{kovacevic_2026ApJ..1006..130K} detected Sy2 variability beyond a constant in the long timescale ASAS-SN data, potentially from the scattered optical emission in Sy2s. Although these studies were conducted at optical wavelengths, they are broadly consistent with our MIR results, in which Sy1s exhibit greater variability in both W1 and W2 than Sy2s do.

\begin{figure}[ht]
\centering
\includegraphics[width=0.45\textwidth,trim={0cm 0cm 0cm 0cm}]{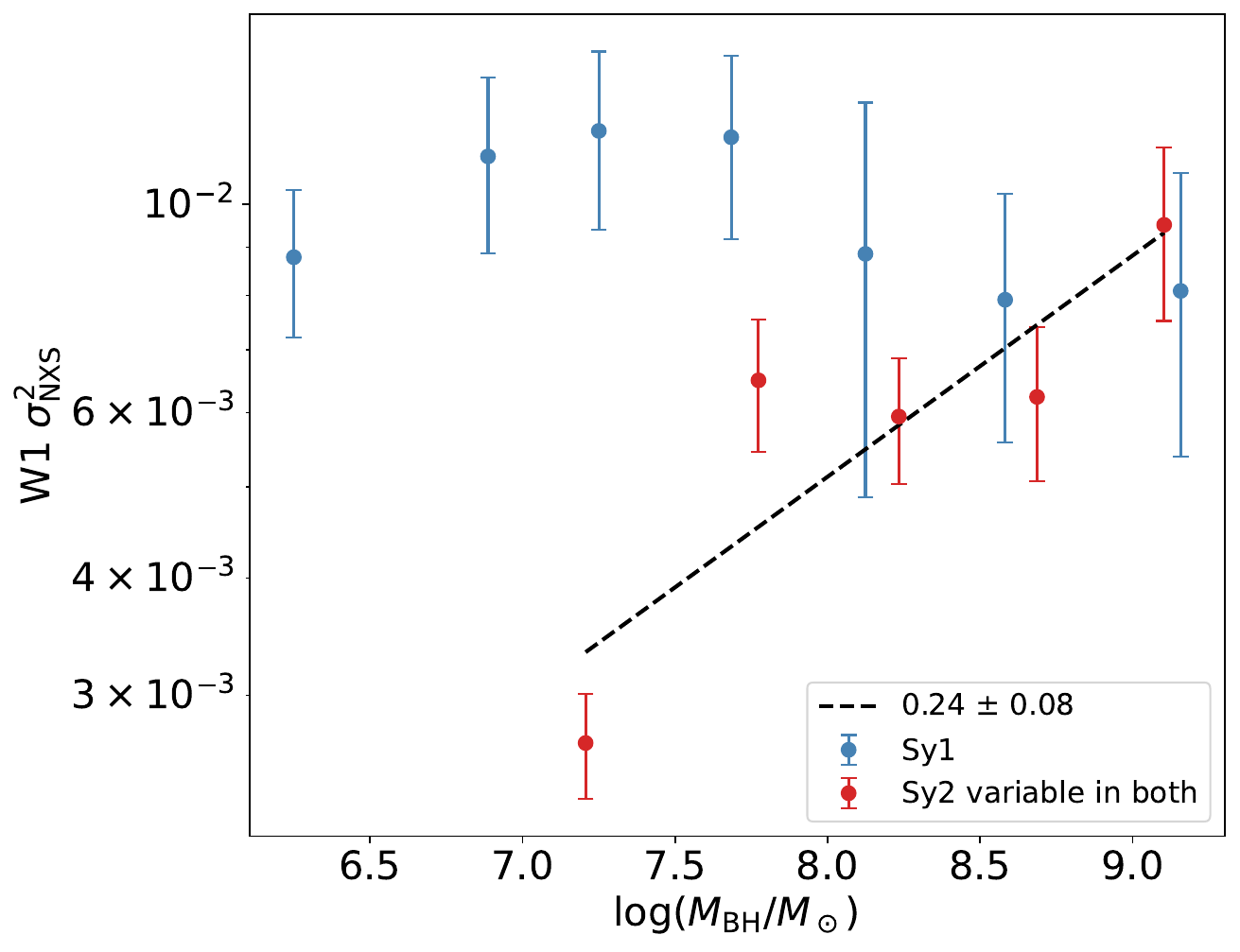}
\includegraphics[width=0.45\textwidth,trim={0cm 0cm 0cm 0cm}]{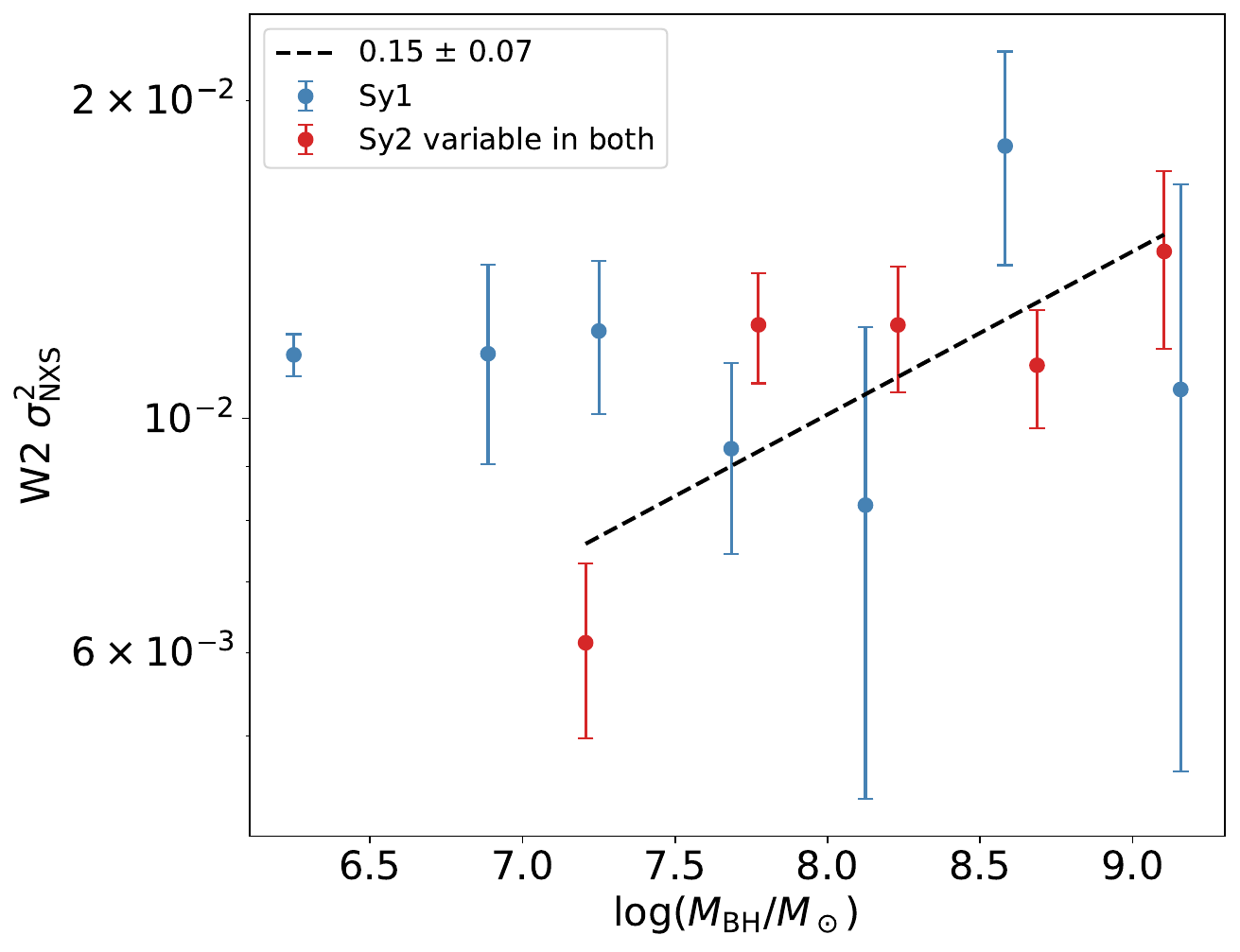}
\caption{Relationship between $\sigma_{\textrm{\tiny{NXS}}}^2$ and black hole mass in the W1 (left) and W2 (right) bands. Fitting the variable Sy2 sample with a power law reveals a positive relationship with slopes of 0.24 $\pm$ 0.08 and 0.15 $\pm$ 0.07, respectively} \label{Mbh}
\end{figure}

\begin{figure}[ht]
\centering
\includegraphics[width=0.45\textwidth,trim={0cm 0cm 0cm 0cm}]{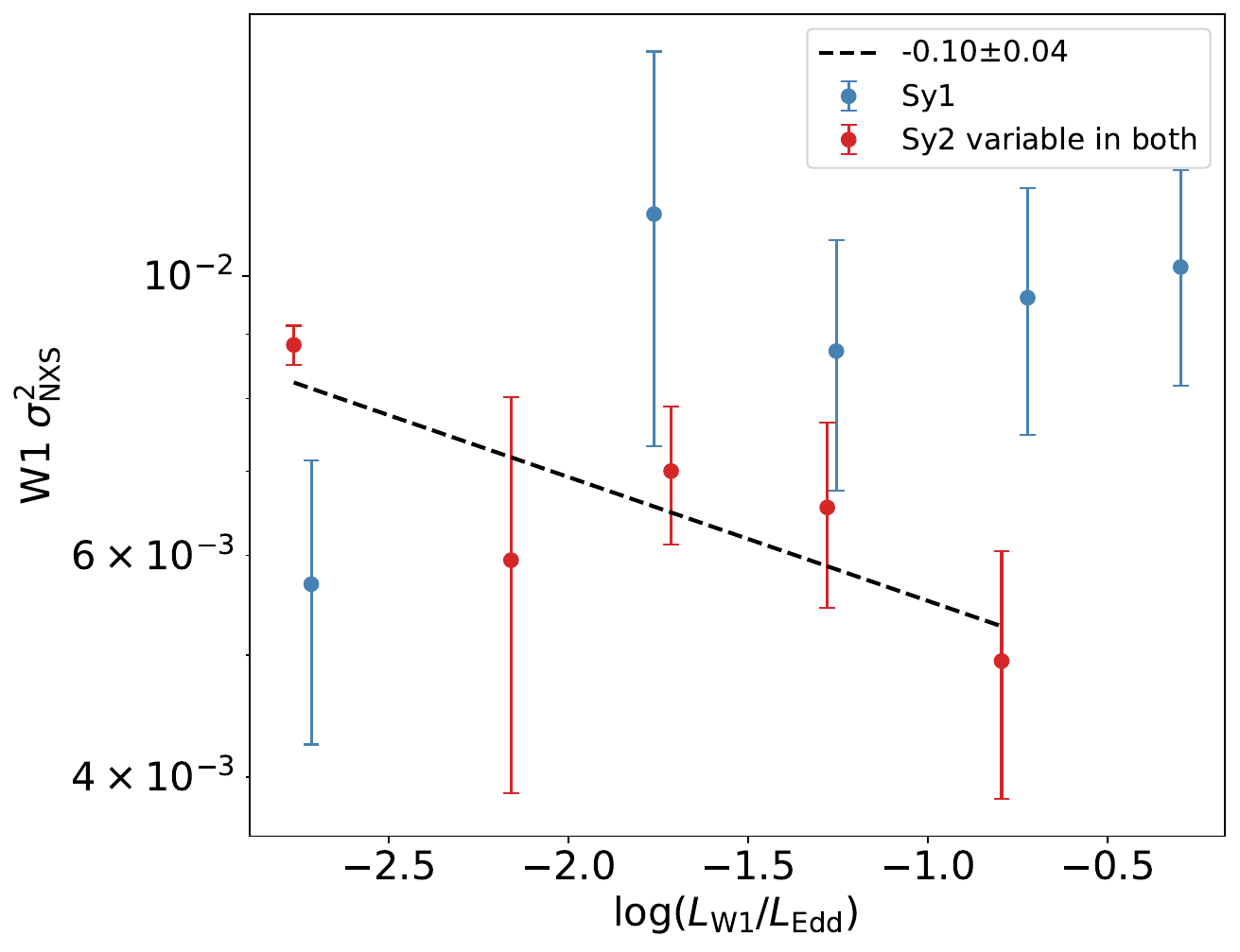}
\includegraphics[width=0.45\textwidth,trim={0cm 0cm 0cm 0cm}]{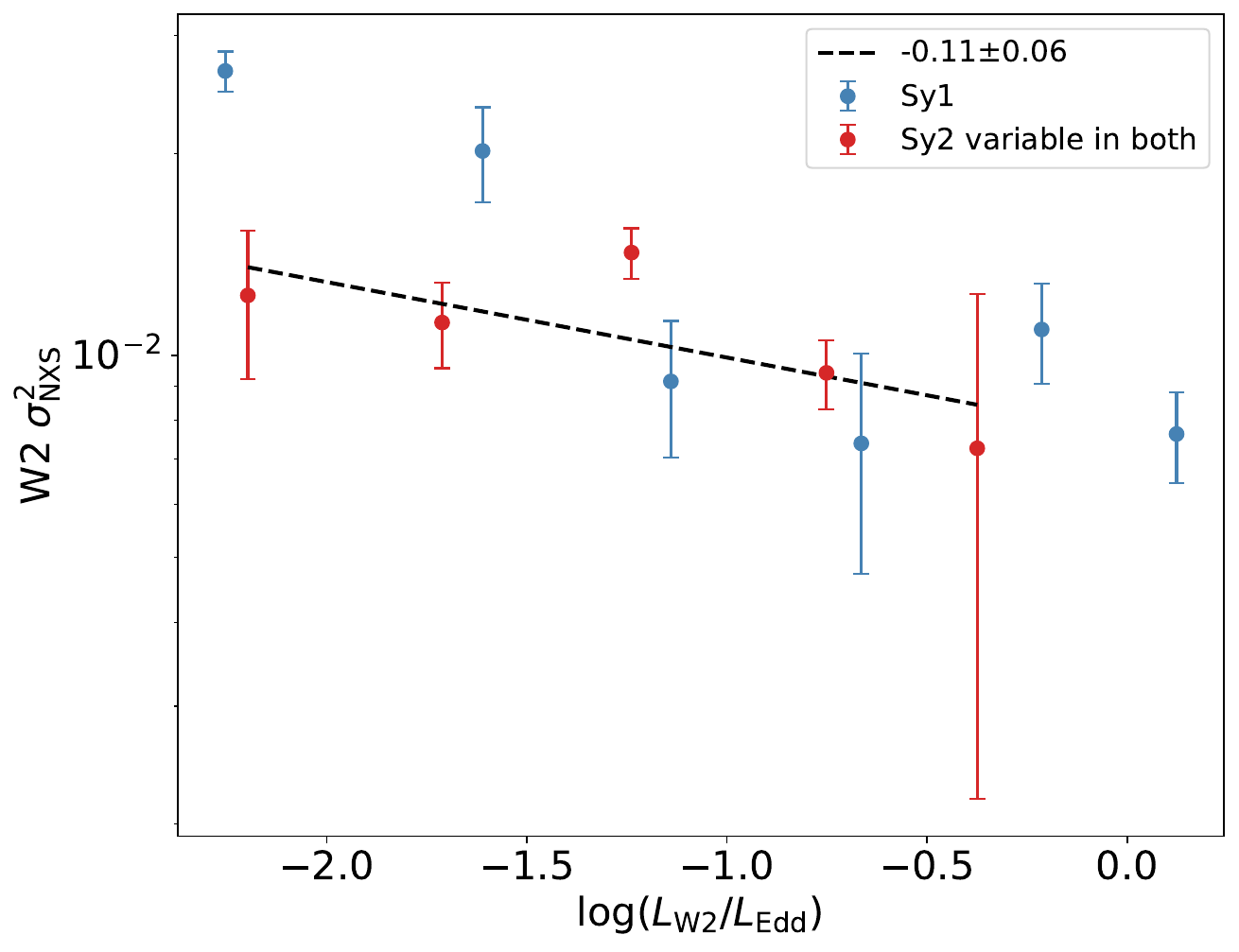}
\caption{Relationship between $\sigma_{\textrm{\tiny{NXS}}}^2$ and a modified eddington ratio in the W1 (left) and W2 (right) bands. Fitting the variable Sy2 sample with a power law reveals a inverse relationship with slopes of -0.10 $\pm$ 0.04 and -0.11 $\pm$ 0.06, respectively} \label{edd}
\end{figure}

Further, we examined how MIR variability relates to different AGN parameters. First, we investigated the relationship between variability and black hole mass, using black hole mass measurements from the BASS Data Release 2 \citep{koss_2022ApJS..261....2K}. By dividing the data into equal-width linear bins and using the standard error of the mean as the error estimate, we found a visually apparent, positive trend between variability and black hole mass for the variable Sy2 sample in both the W1 and W2 bands (Figure~\ref{Mbh}). Performing a correlation test between the normalized excess variance and the black hole mass on the non-binned data we find a weak positive correlation with black hole mass. Power law fits yield slopes of 0.24$\pm$0.08 and 0.15$\pm$0.07 for W1 and W2, respectively. This trend is not observed in our Sy1 population; however, the number of sources with available black hole mass measurements is substantially larger for the Sy2 population in this sample thanks to the stellar-mass black hole mass correlation available for Sy2s. The positive variability amplitude against black hole mass relation is consistent with previous findings (\citealt{Macleod_2010ApJ...721.1014M,kovacevic_2025ApJ...985..177K}) that optical variability scales positively with black hole mass in a sample of Sy1s. Using the available black hole mass measurements, we calculated a modified Eddington ratio based on the W1 and W2 luminosities relative to the Eddington luminosity. As shown in Figure~\ref{edd}, a decreasing trend between variability and Eddington ratio is visually apparent in the W1 band for the variable Sy2 sample. Performing another correlation test between the normalized excess variance and modified eddington ratio on the non-binned data yielded a weak negative correlation with the eddington ratio. Power law fits yield slopes of $-$0.10$\pm$0.04 and $-$0.11$\pm$0.06 for W1 and W2, respectively, indicating that the inverse relationship is present in both MIR bands for the variable Sy2 sample. This anti-correlation is also consistent with the findings of \citet{Macleod_2010ApJ...721.1014M} and \citet{kovacevic_2025ApJ...985..177K}, who found a similar inverse relationship between optical variability and Eddington ratio for a sample of Sy1s.

\begin{figure}[ht]
\centering
\includegraphics[width=0.405\textwidth,trim={0cm 0cm 0cm 0cm}]{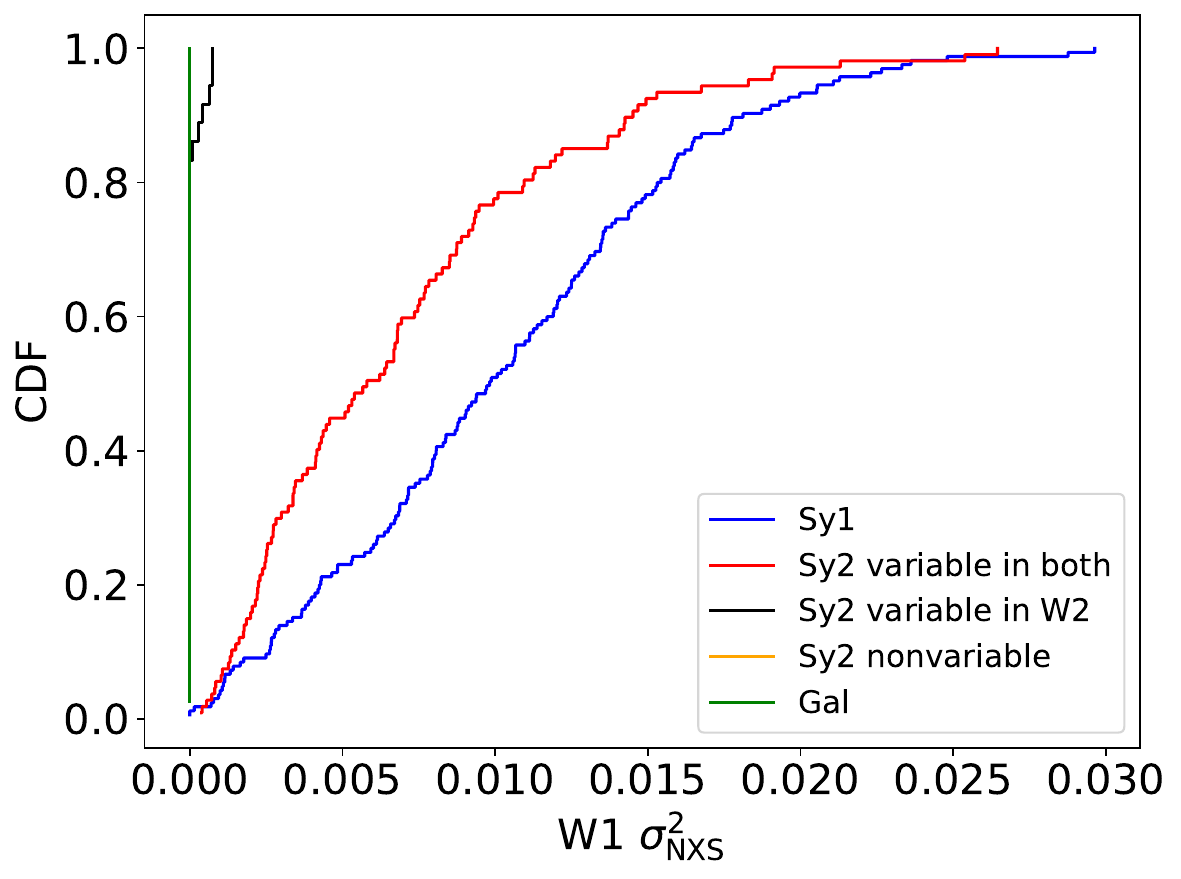}
\includegraphics[width=0.4\textwidth, trim={0cm 0cm 0cm 0cm}]{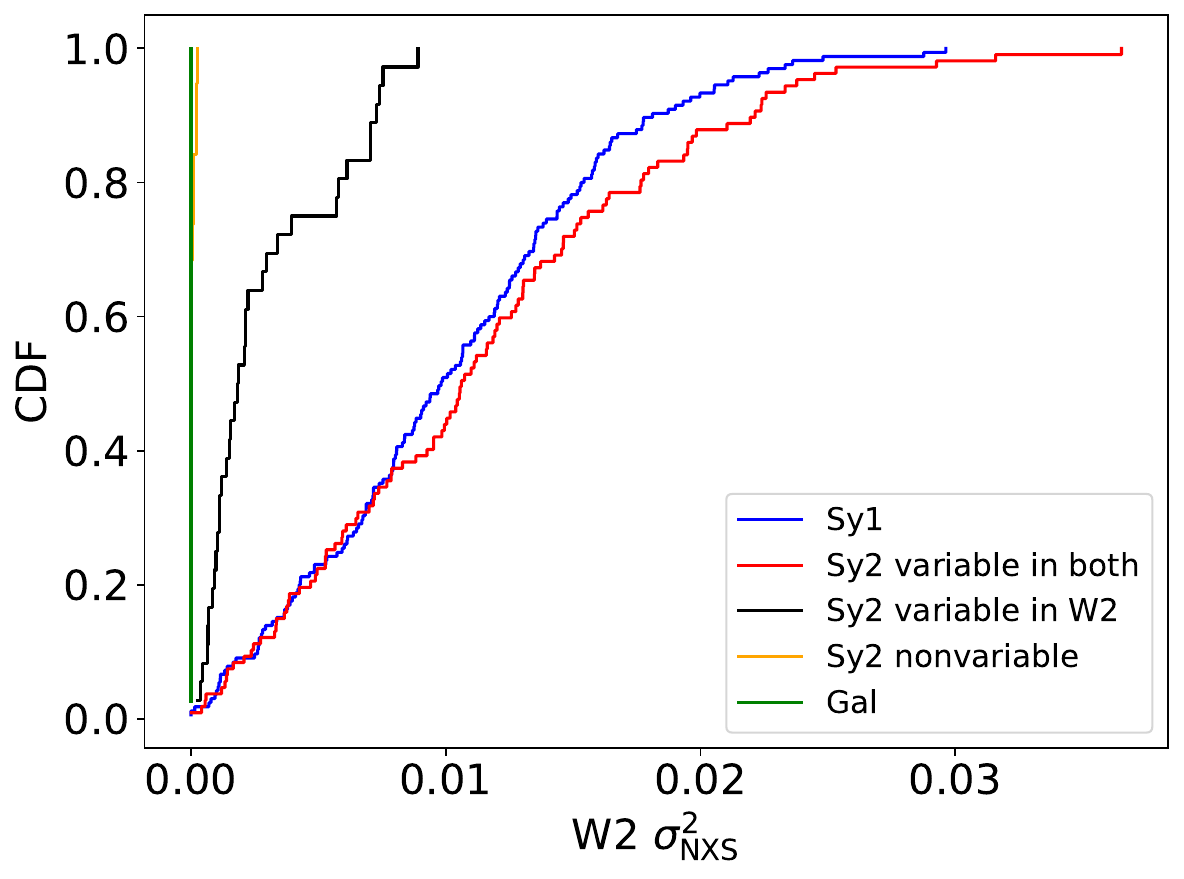}
\caption{The cumulative distribution function of the excess variance for W1 and W2 of the Sy1 and Sy2 samples. The Sy1 and Sy2 variable in both excess variance distributions are significantly different in W1, with a KS-test p-value of $3.3\times10^{-5}$. In the W2 band, the two distributions are less distinguishable, with a p-value of $0.04$.}
\label{CDF_wise}
\end{figure}

\subsection{Optical vs Mid-Infrared Variability Amplitude}

\begin{figure}[ht]
\centering
\includegraphics[width=0.4\textwidth,trim={0cm 0cm 0cm 0cm}]{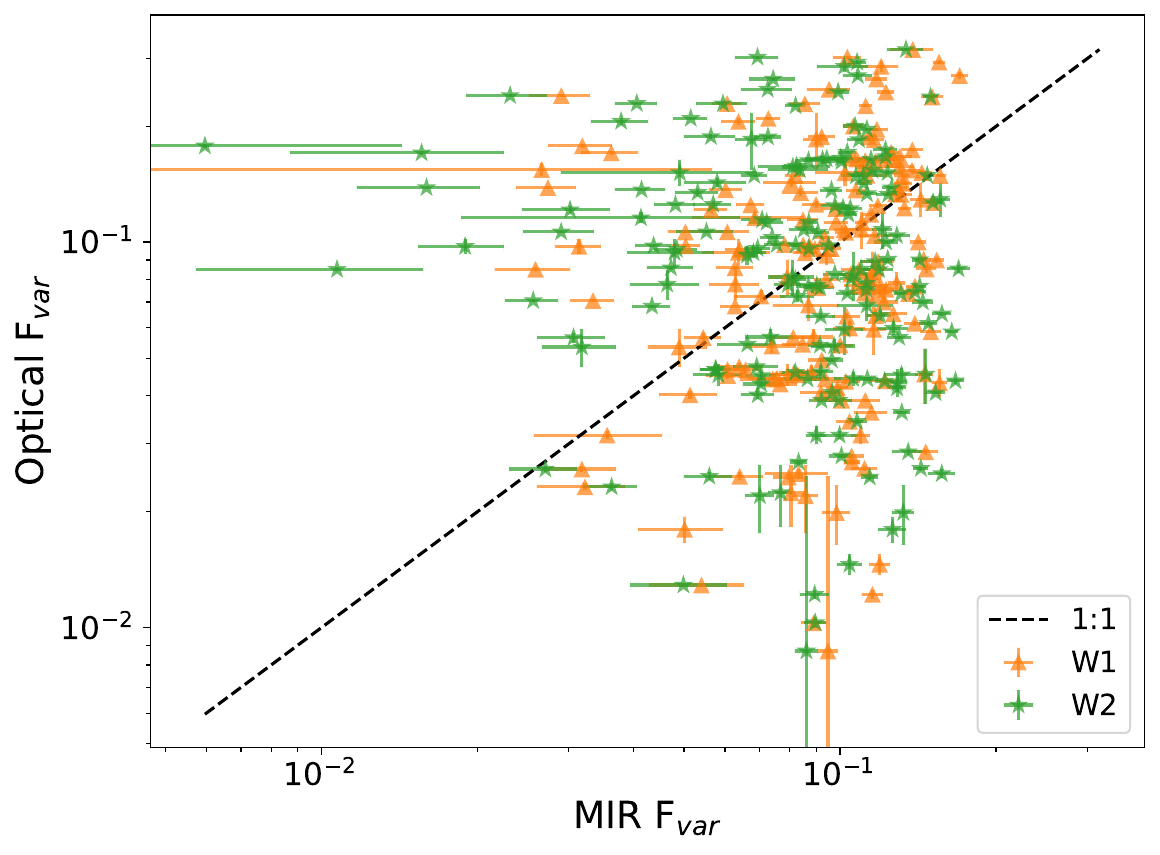}
\includegraphics[width=0.4\textwidth,trim={0cm 0cm 0cm 0cm}]{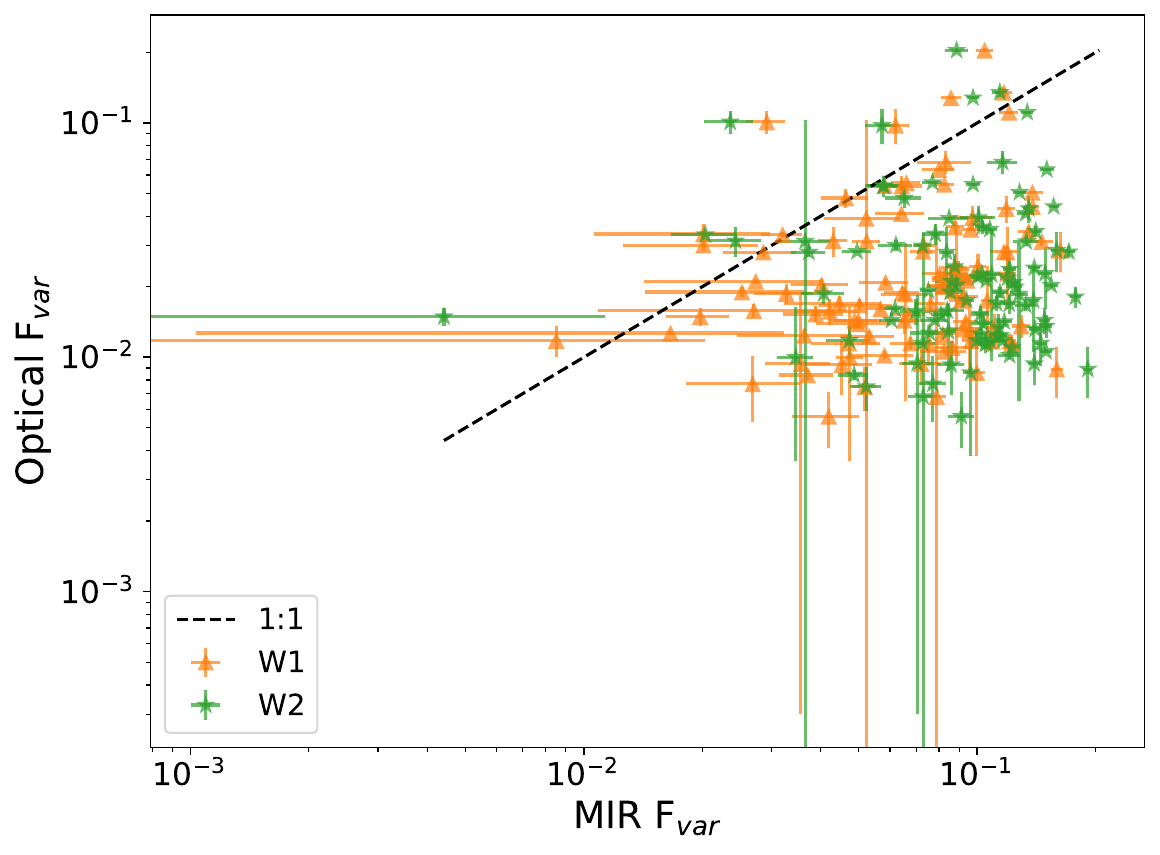}
\caption{left: Scatter plot visually representing how individual sources variability changes based on wavelength for our Sy1 sample. With the large scatter presented a relationship between fractional variabilty and wavelength is undetermined. right: Same scatter plot for Sy2 sources. Most sources fall below the 1:1 line indicating that Sy2 sources tend to be more variable in the MIR.} \label{scatter}
\end{figure}

We used the fractional variability to compare the variability amplitudes between optical and MIR bands.
Figure~\ref{scatter} shows the scatter plots of fractional variability in optical and MIR for individual Sy1 and Sy2 sources.
For the Sy1 population there is a large scatter within the data  and no clear dependence between the fractional variability in the optical and MIR bands can be determined.
A portion of Sy1s exhibit larger optical variability than the MIR band, and yet a significant larger fraction of Sy1s showing the opposite scenario with larger MIR variability.
However for the Sy2 population almost all the sources are below the 1:1 line signifying that Sy2s tend to be more variable in the MIR bands than the optical band. 

\begin{figure}[ht]
\centering
\includegraphics[width=0.4\textwidth,trim={0cm 0cm 0cm 0cm}]{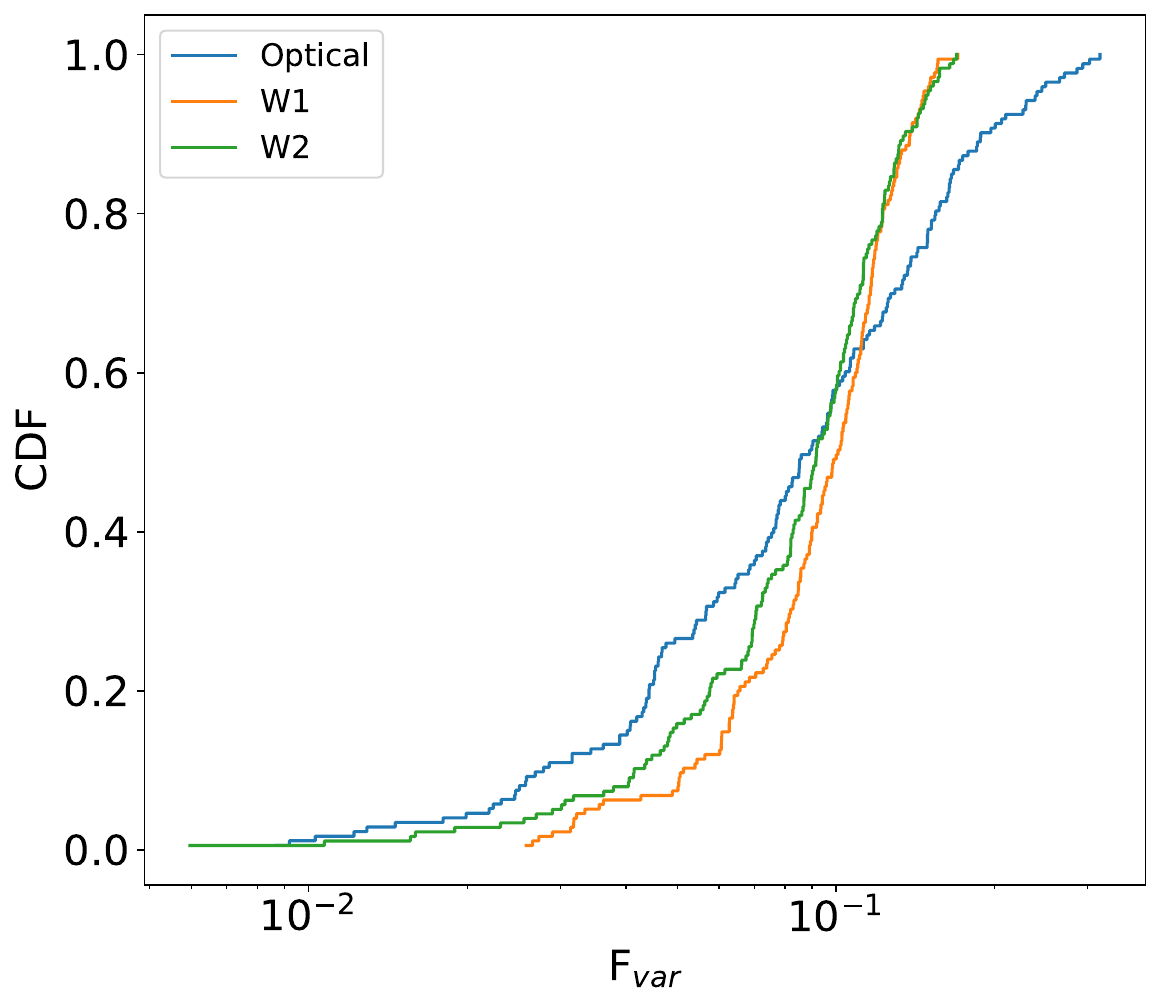}
\includegraphics[width=0.4\textwidth,trim={0cm 0cm 0cm 0cm}]{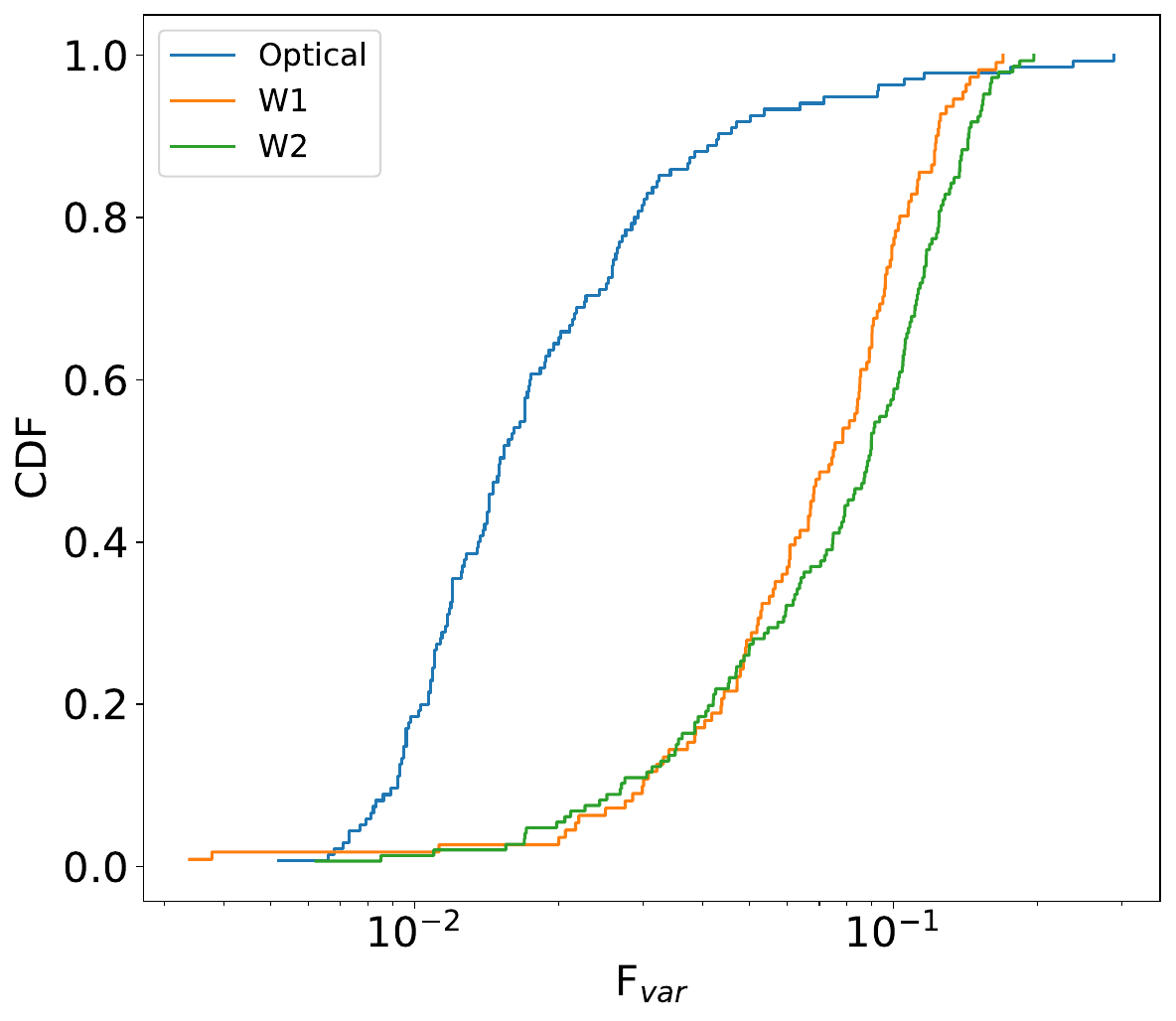}
\caption{left: The cumulative distribution function of the Fractional Variability of Sy1s in the Optical ASAS-SN light curves compared to the NEOWISE W1 and W2 light curves showing little to small differences in fractional variability between the three bands. Right: Same as the left panel but for our Sy2s visually displaying an increase in fractional variability with an increase in wavelength. KS-test p-values between the optical and W1, optical and W2, and W1 and W2 are 1.6$\times 10^{-16}$, 4.2$\times 10^{-31}$, and 1.4$\times 10^{-4}$.} \label{KS_optical}
\end{figure}

KS-tests were performed on the $F_{\textrm{var}}$ of the optical and MIR (W1 and W2) light curves from our Sy1 and entire Sy2 sample. The comparison between the optical and W1 bands for Sy1s yielded a p-value of $1.2\times10^{-11}$, the result between optical and W2 band for Sy1s yielded a p-value of $1.7\times10^{-7}$. Both results are well below the null hypothesis threshold, suggesting that Sy1s have different optical and MIR variability.
The mean fractional variability values show that
Sy1 optical variability tends to be just below that of the MIR with an average of $F_{var}$, optical, Sy1= $0.066 \pm 0.001$ compared to MIR with  $F_{var}$, W1, Sy1 = $0.103 \pm 0.005$ and $F_{var}$, W2, Sy1 = $0.095 \pm 0.005$ for W1 and W2 respectively. Combined with the visual representation and the comparison between the optical and MIR fractional variability shown in the left panel of Figure~\ref{KS_optical} after performing a correlation test, no clear correlation between optical and MIR fractional variability was determined within the Sy1 sample.
This is different when we compare the optical and MIR of our entire Sy2 sample, in the right panel of Figure~\ref{KS_optical} it is clear Sy2 variability increases with an increase in wavelength where optical wavelengths displays low variability and the W2 band the highest variability. Performing KS-tests on our entire Sy2 sample between optical and W1, optical and W2 and W1 and W2, yields p-values of 1.6$\times 10^{-16}$, 4.2$\times 10^{-31}$, and 1.4$\times 10^{-4}$ respectively, with an average fractional variability of $F_{var}$, optical, Sy2 = $0.015 \pm 0.001$, $F_{var}$, W1, Sy2 = $0.050 \pm 0.003$ and $F_{var}$, W2, Sy2 = $0.081 \pm 0.005$ for the optical, W1 and W2 band. However performing a correlation test revealed no clear correlation between the optical and MIR fractional variability either. This is consistent with a scenario in which torus variability dominates in Sy2s, \citet{cerone_2026A&A...709A.245C}, found that in the NIR torus dominated Sy2s follow the trend of exhibiting higher levels of variability towards redder wavelenghts.  Our analysis extends this trend to the MIR band.
For a small fraction of Sy2s with large fractional variability and higher optical variability, they are potentially misclassified or changing-look AGN.

\subsection{Mid-Infrared Optical lags}

\begin{figure}[ht]
\centering
\includegraphics[width=0.4\textwidth,trim={6cm 0cm 6cm 0cm}]{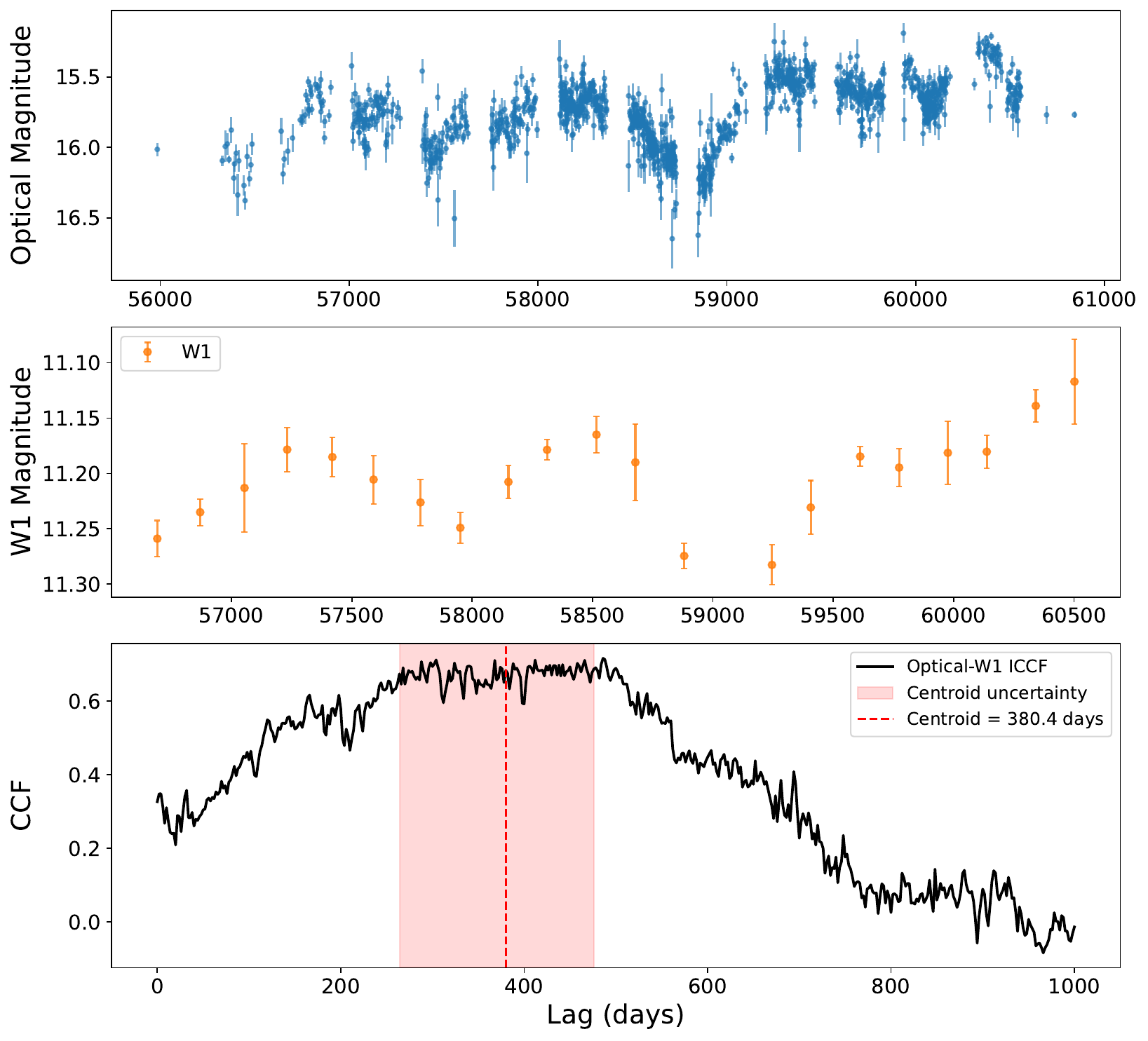}
\caption{Top and Middle: ASAS-SN and W1 NEOWISE light curves for the Sy1 2MASXJ14364961-1613410 revealing an echo between the optical and MIR variability. Bottom: The result of performing a cross-correlation function on the optical and Mid-IR light curves, estimating the MIR time delay. The red dashed line representing the location of the most probable lag time with uncertainties represented by the shaded regions. }\label{lag}
\end{figure}

In accordance with the AGN unification model, UV/optical light traces emission from the accretion disk, while infrared light mainly traces thermal re-emission from the dusty torus  after it absorbs the UV/optical photons with an additional disk emission component. It has been shown that there is often an ``echo'' effect between the two signals reaching the observer. This time lag ($\tau$) is often cited in dust reverberation mapping studies (\citealt{blandford_1982ApJ...255..419B,cackett_2021iSci...24j2557C,Izviekova_2026arXiv260506517I}), where it can be used to probe various properties of the accretion disk. This lag can be calculated by performing a cross-correlation function on the optical and infrared light curves of AGNs and determining where the peak lag is located. 
We performed an interpolated cross correlation function (ICCF) (\citealt{perterson_1998PASP..110..660P,sun_2018ascl.soft05032S}) on the variable Sy1s in our sample to determine the time delay between the variable optical and MIR light curves. The ICCF was chosen to combat the unevenly sampled ASAS-SN and sparse nature of NEOWISE light curves.
From our sample of 165 Sy1 light curves, we were able to calculate quality time lags for 93 sources, with lag times of W1-optical and W2-optical ranging from 22--735 and 28--786 days with a median lag time of 181 days and 259 days respectively. Figure~\ref{lag} shows one example, displaying both the optical and W1 light curves of the Sy1 2MASXJ14364961-1613410 along with the result of the cross-correlation. A positive $\tau$ indicates the MIR light lagging behind the optical light. Visually, this is seen in the light curves as a peak in the optical being followed by a similar peak in the MIR.

\begin{figure}[ht]
\centering
\includegraphics[width=0.3\textwidth,trim={0cm 0cm 0cm 0cm}]{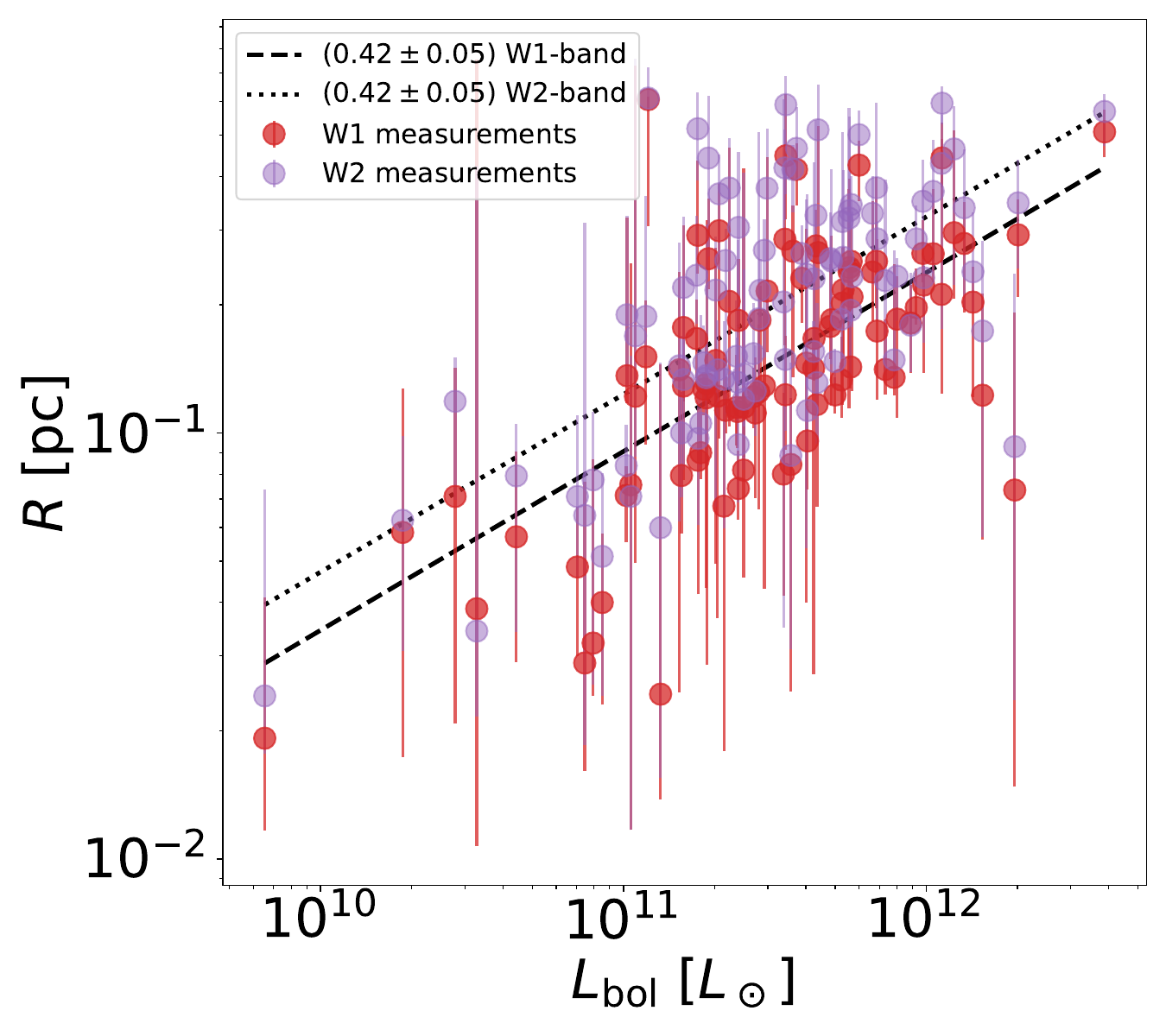}
\includegraphics[width=0.3\textwidth,trim={0cm 0cm 0cm 0cm}]{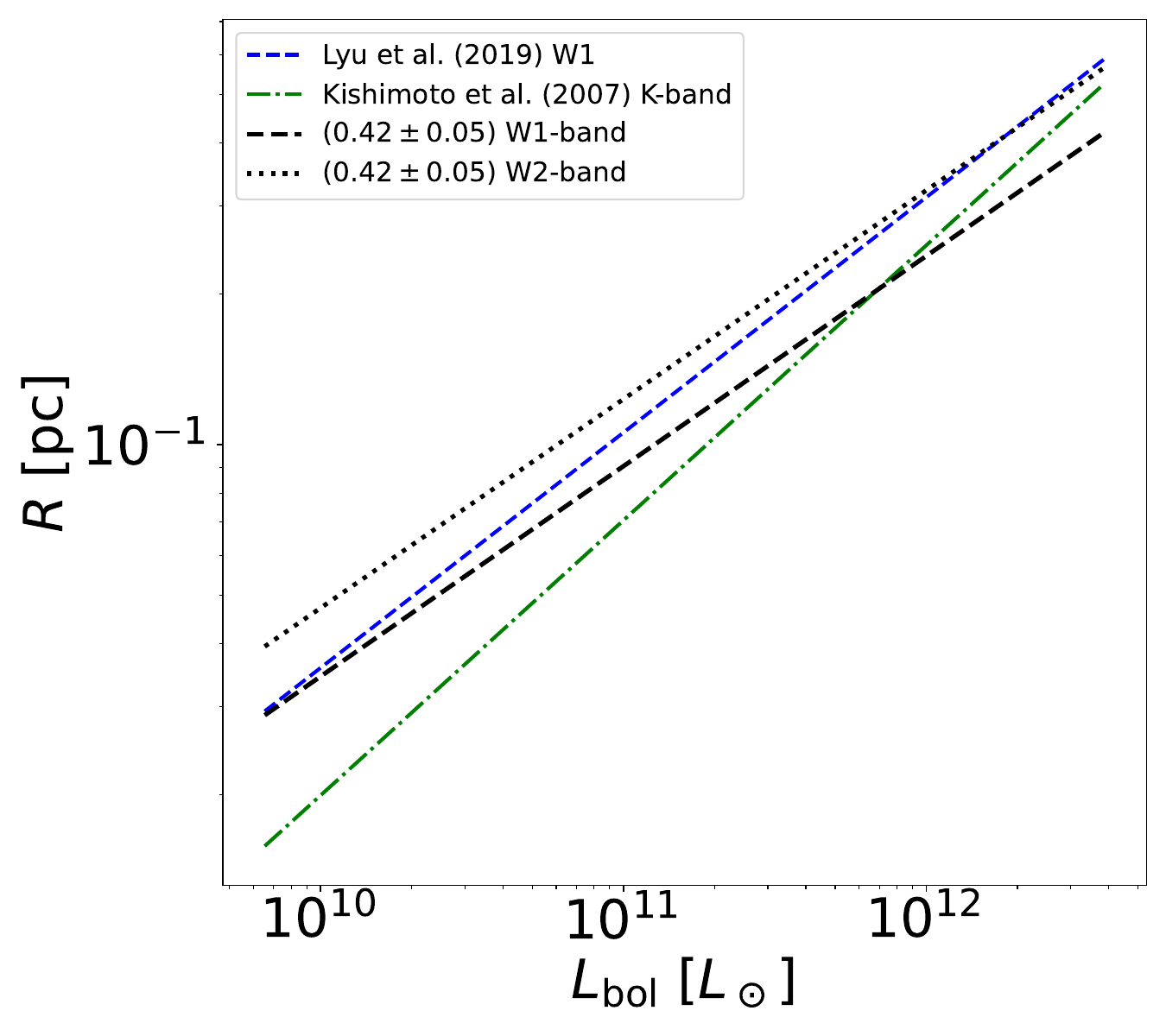}
\includegraphics[width=0.3\textwidth,trim={0cm 0cm 0cm 0cm}]{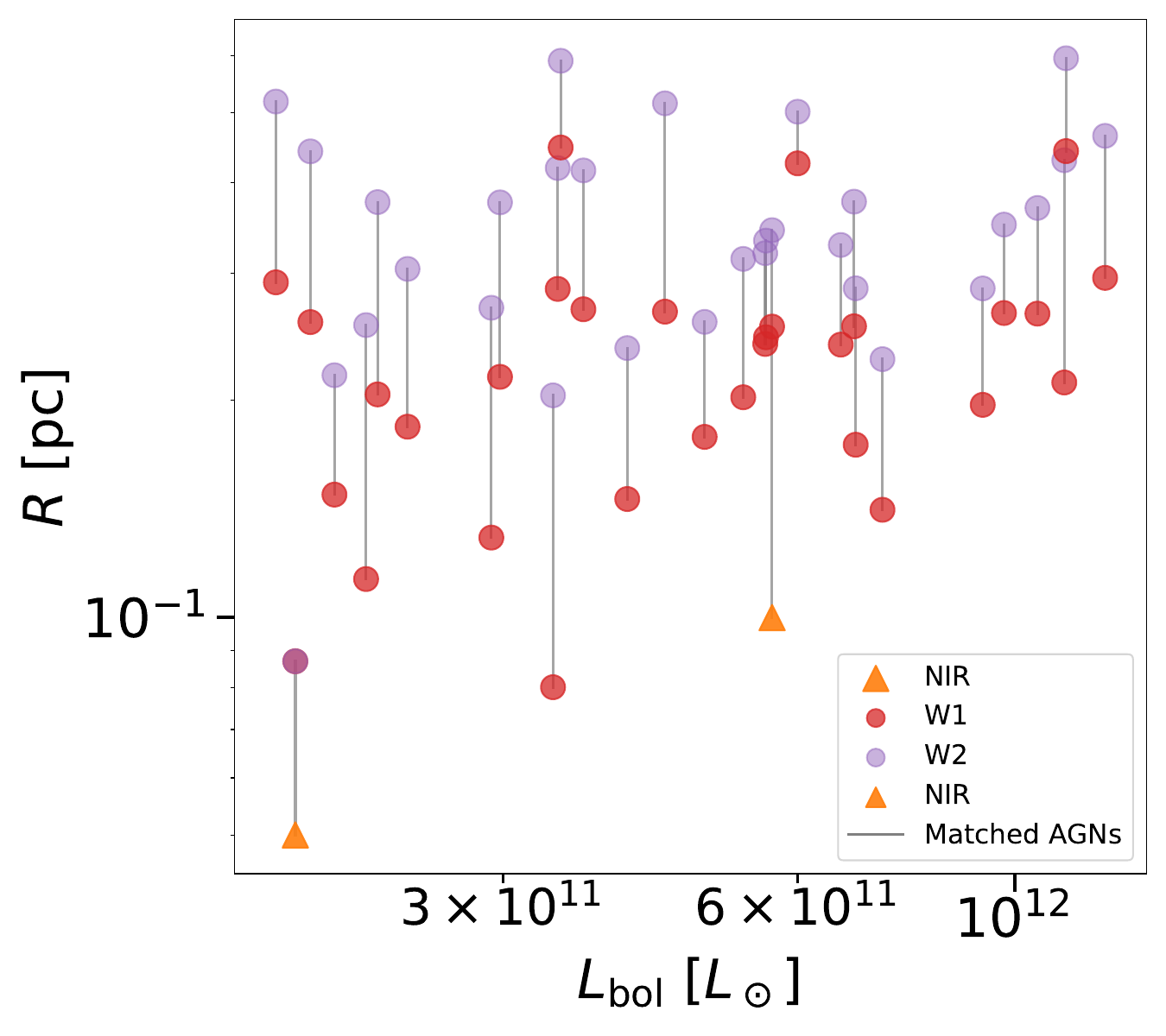}
\caption{Left: The relationship between bolometric luminosity and estimated dust radii traced by both W1 and W2 with our luminosity-radius relations for both W1 and W2. Middle: Same as left but compared with relations found in \cite{Kishimoto_2007A&A...476..713K} in $K$-band and \cite{lyu_2019ApJ...886...33L} in W1. Right: Individual sources with non-overlapping W1 and W2 radii including two sources NGC 5548 and Mrk 509 with $K$-band measurements represented as orange triangles \citep{Minezaki_2019}}\label{torus}
\end{figure}

The lag time measurement is utilized to estimate the dust radius for the 93 Sy1s using Equation \ref{equation4} below where $\tau$ has been corrected to include cosmological time dilation. The results are displayed in the left panel of Figure~\ref{torus}, showing the relationship between the bolometric luminosity and the dust radius, where our estimated radii range from 0.02 -- 0.61 pc.

\begin{equation}\label{equation4}
R = \tau_{\textrm{rest}} \times c
\end{equation}

Performing power-law fits to the W1 and W2 measurements, $R \propto L^{\beta}$, we derived slopes of $\beta = 0.42\pm0.05$ for both W1 and W2.
We also found on average larger radii in the W2 band than those in W1 band. 
Of the 93 sources with reliable
lag measurements, 62 have consistent radii between the W1 and W2 bands, potentially caused by the sparse sampling of NEOWISE light curves. Using the remaining 31 Sy1s, we fit a relation between W1 and W2 sizes as $R_{W2} = 1.2R_{W1}+0.02$.
 Additionally, two sources in our sample, NGC 5548 and Mrk 509, have near-infrared $K$-band radius measurements from \citet{Minezaki_2019}. These sources are shown in the right panel of Figure~\ref{torus}. The measured radii increase toward longer wavelengths, with the relative cooler dust traced by W2 located at larger radii than the warmer dust traced by W1 and the hottest dust traced by the $K$-band.
We compare these relations with the luminosity–radius relation described by \citet{Kishimoto_2007A&A...476..713K}, derived from near-infrared $K$-band dust reverberation mapping, and the relation from \citet{lyu_2019ApJ...886...33L}, which was derived using NEOWISE W1 measurements (middle panel of Figure~\ref{torus}). Our relations are broadly consistent with those found in the literature, considering the large scatter and systematics in each set of measurements. 

\begin{figure}[ht]
\centering
\includegraphics[width=0.3\textwidth,trim={0cm 0cm 0cm 0cm}]{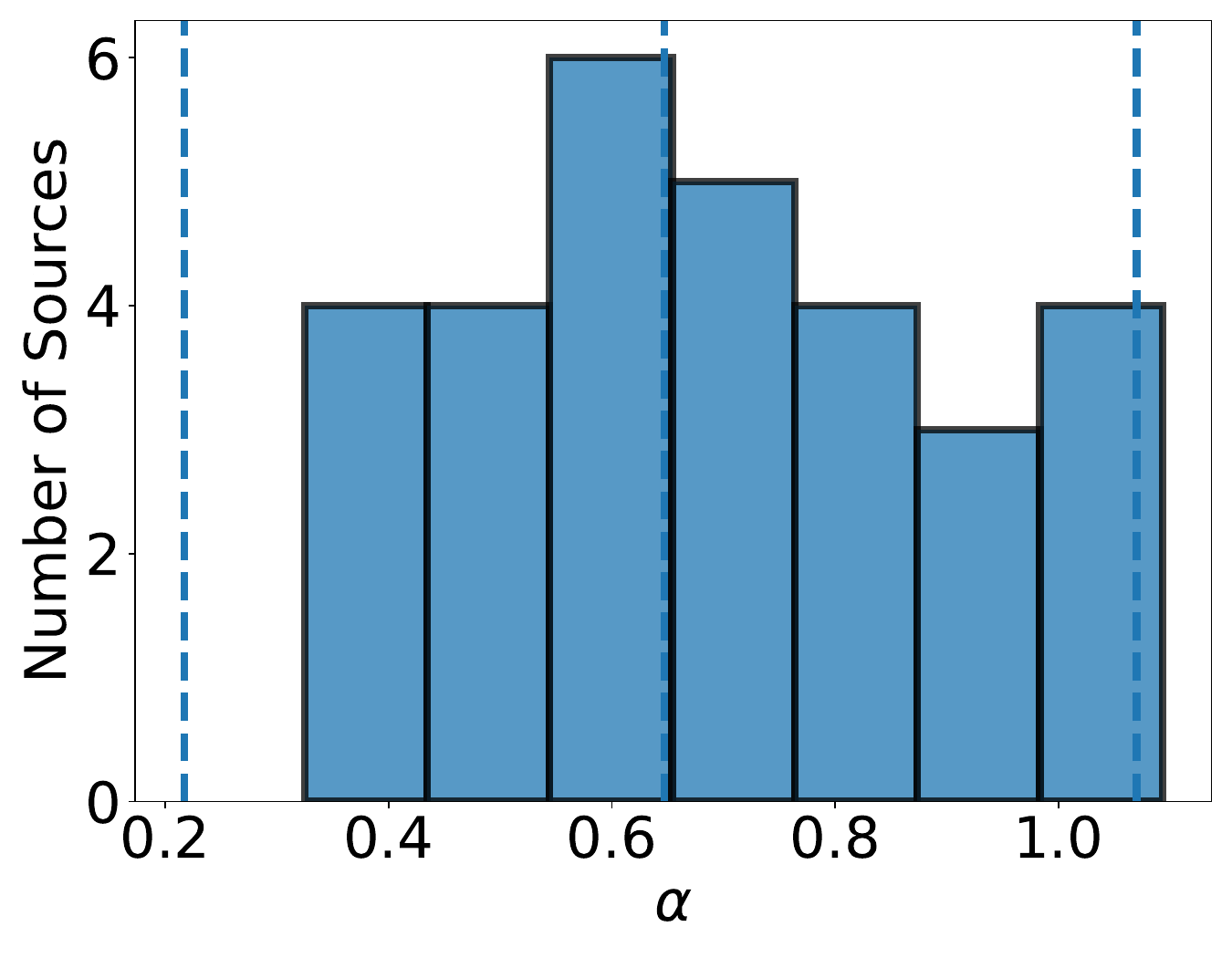}
\caption{distribution of $\alpha$ values estimated from fitting individual sources with non-overlapping W1 and W2 radii. For NGC 5548 and Mrk 509 their $K$-band measurements were included in the fit. The vertical dotted line is the estimated $\alpha$ measurement with uncertainties from utilizing our W1 and W2 relations.}\label{alpha}
\end{figure}

We can estimate the temperature profile of the torus by assuming $T\propto R^{-\alpha}$ or equivalently $R_\lambda \propto \lambda^{1/\alpha}$, from the multiple size measurements of the same source in different bands. 
It can also be applied between fitting relations of torus size at different bands, where we used our own W1 and W2 size relations, but not including literature relations due to systematics between different measurements. 
We derive $\alpha$ values for each source with non-overlapping radii, including NGC 5548 and Mrk 509 using their near-infrared measurements. The resulting $\alpha$ values range from 0.3--1.1. Using the W1 and W2 radius–luminosity relations and a representative bolometric luminosity of $5.0\times10^{11} L_\odot$, we derive $\alpha=0.65$. The distribution of the individual $\alpha$ values is shown in Figure~\ref{alpha}, with the vertical line indicating the $\alpha$ value derived from our radius–luminosity relations.

\section{Discussion}

In our analysis of the normalized excess variance revealed that almost all Sy1s in our sample are variable and the majority of the Sy2s show measurable variability amplitudes, with higher values presented in the W2 band. Although Sy1s generally exhibit larger variability amplitudes, based on the K-S tests performed on both the W1 and W2 bands of our Sy1 and Sy2 samples, we find that their variability in the MIR are not significantly distinguished in W2 with a p-value of 0.04. This is consistent with the study by \citet{papadakis2024x}, they reported similar power spectral density (PSD) shapes for Sy1s and Sy2s in the X-ray band suggesting that the underlying variability process is similar in both populations. 
One explanation for this similar variability behavior is that in the W1 band the MIR variability is mostly dominated by the dust torus, however since W1 traces the hot dust located at a closer distance to the central engine and we have a clear line-of-sight to the central engine of Sy1s variability from the disk can contribute to the larger normalized excess variance values presented in the W1 band. In contrast as we move to longer wavelengths the effects of orientation on MIR variability diminish. At the larger radii predominantly traced by W2, the emitting dust is extensively from the torus. This makes the MIR band a valuable channel to study AGN variability and perform variability selection, with respect to obscured and unobscured AGNs. We also analyzed the relationship between AGN properties (black hole mass and eddinton ratio) and variability for the Sy2s that exhibited variability in both MIR bands. Performing a correlation test we found that the variability has a weak statistically significant positive correlation to the black hole mass and a weak statistically significant negative correlation to the Eddington ratio. This result is consistent with similar studies performed in the optical band (\citealt{kovacevic_2025ApJ...985..177K,Macleod_2010ApJ...721.1014M}), that Sy1 variability follows this same trend. 

A comparison of the NEOWISE W1 and W2 bands with the ASAS-SN optical band reveals that, for the Sy1 population, there is little difference in the variability amplitudes between the optical and MIR bands. However, there is substantial scatter around the 1:1 line, suggesting that individual sources may have different relative contributions from the accretion disk and torus, with some sources being more torus-dominated while others appear to be more accretion-disk-dominated. In contrast, the Sy2 population exhibits a wavelength-dependent trend, with the variability amplitude increasing toward longer wavelengths, such that most Sy2s fall below the 1:1 line when comparing the optical and MIR bands. 
One possible explanation for this apparent wavelength dependence is that the MIR emission in Sy2s is more strongly dominated by the dusty torus \citep{cerone_2026A&A...709A.245C}. The fraction of non-variable Sy2s in our sample is 0.11 in W2, these sources may experience higher levels of obscuration toward the inner regions of the torus, even suppressing the observed MIR emission and shifting the reprocessed emission toward longer, redder wavelengths, potentially to the far-infrared.

As the warm dust in the torus responds to variations in the accretion disk, the optical–MIR rest frame time delay can be used to estimate the size of the inner dust emitting region of the torus. Using dust reverberation mapping techniques, we estimated the characteristic emitting region radius for 93 Sy1s, obtaining a median time delay of $\tau = 181$ days in W1 and $\tau = 259$ days in W2. From these time delays we calculated the warm dust radii traced by W1 to range from 0.02 -- 0.6 pc, while the cooler dust traced by W2 is found at radii ranging from 0.02 -- 0.61 pc. For individual sources it is noted that the W2 measurements is always located at a larger radii compared to their W1 measurement. The $K$-band is a NIR band that traces the hottest dust close to the sublimation radii, which we expect to measure smaller radii. However, from comparing our W1 and W2 results with the $K$-band dust reverberation mapping results from the literature we found that our W1 sizes are not significantly larger, possibly due to a difference in systematics in the data. Two sources in our sample NGC 5548 and Mrk 509 have measured $K$-band radii \citep{Minezaki_2019} that are smaller than the radii of W1, as expected. We fit our W1 and W2 measurements with power-law fits and compared our relations, $R \propto L^{0.42\pm0.05}$, with relations from the literature including \citet{Kishimoto_2007A&A...476..713K} probing the radius-luminosity relationship in the $K$-band and relations found in \citet{lyu_2019ApJ...886...33L} for the W1 and W2 band. Our relations are in agreement with the relations discussed in \citet{lyu_2019ApJ...886...33L} with their W1 and W2 relation falling between ours. 
This generally confirms the basic reprocessing picture of the torus, where for a similar dust temperature in the torus, the incident flux, $L/R^2$, is roughly a constant.
The relation described in \citet{Kishimoto_2007A&A...476..713K} predicts comparable radii to our W1 relation, one explanation for this comparable result is due to a difference in systematics presented in the data. From our relations and individual sources, previously unexplored in the literature, we estimated tori temperature profiles $\alpha$ from 0.3--1.1 with an estimated $\alpha=0.65$ derived from our W1 and W2 relations. For a standard geometrically thin, optically thick accretion disk, the radial temperature profile is expected to follow $T_{\rm eff}(R)\propto R^{-0.75}$ \citep{shakura_1973A&A....24..337S}. In the slim disk regime, the temperature profile becomes shallower, following approximately $T_{\rm eff}(R)\propto R^{-0.5}$ \citep{Wang_1999}. For the torus dust temperature profile, a pure blackbody model balancing from heating from the accretion disk predicts $T \propto R^{-0.5}$ for a smooth distribution of dust. \citet{Barvainis_1987ApJ...320..537B} considered additional absorption effect, yielding a shallow temperature gradient of $T \propto R^{-0.36}$. In clumpy torus models, where dust grains are modeled as clouds, \citep[e.g.,][]{hoing_2010A&A...523A..27H,leftley_2024A&A...686A.204L} a broad temperature distribution from dust clouds at a given radius are predicted. 

\citet{hoing_2013ApJ...771...87H} modeled the infrared emission and interferometric data of NGC 3783 using two components representing the hot NIR and warm MIR emission. Their models yielded temperature gradients corresponding to 0.28 and 0.26. Similarly, \citet{burtscher_2009ApJ...705L..53B} modeled the 8--13 $\mu$m emission of NGC 4151 using graybody models with a smooth power-law dependence and found acceptable fits for temperature gradients corresponding to $0.35<\alpha<0.6$. Based on our temperature gradients and what has been reported in the literature for the accretion disk and few sources for the dust tori, we interpret that our sources are broadly consistent with predictions from different torus models and better measurements are needed to better constrain the temperature profile. 

\section{Conclusions}
Finally, we summarize the conclusions of this paper below.
\begin{itemize}
\item Nearly all Sy1s in our sample are classified as variable, while the majority of Sy2s exhibit measurable variability, highlighting that the mid-IR is an important channel for studying AGN variability from the torus. Furthermore, the variability amplitudes of Sy1s and Sy2s are comparable in W2, with both populations having a mean $\sigma^2_{\textrm{\tiny{NXS}}}$ of 0.01 and a K-S test $p$-value of 0.04. This demonstrates that significant mid-IR variability is present across both unobscured and obscured AGN populations, signaling the MIR as a relatively unbiased channel for studying AGN variability and selecting variable AGNs regardless of obscuration.

\item Sy1s remain overall more variable than Sy2s despite comparable MIR variability amplitudes in W2. While the MIR variability is expected to originate predominantly from the dusty torus, additional variability from the accretion disk can contribute to the MIR emission, potentially producing the larger variability amplitudes observed in Sy1s. To investigate the connection between variability and AGN properties, we find a weak statistically significant positive correlation between variability amplitude and black hole mass in Sy2s, and a weak statistically significant negative correlation with Eddington ratio. These trends provide insight into how the physical properties of the central engine influence MIR variability in obscured AGNs.

\item 
Comparing the MIR and optical variability, we find that the Sy2 sample follows a wavelength-dependent trend in which the fractional variability increases toward longer wavelengths, while this wavelength dependence is not observed in the Sy1 sample. The wavelength dependence in Sy2s suggests that their variability may be more coherently associated with emission from the dusty torus. The non-variable Sy2s are expected to experience the strongest obscuration, suppressing variability even in MIR while allowing variability to become more apparent at longer wavelengths, where the obscuring optical depth is lower.
    
\item From multi wavelength observations a cross correlation analysis yielded time lags for 93 Sy1s between either NEOWISE band and optical ASAS-SN band, from which dust radii is estimated ranging from 0.02--0.61 pc. 
We confirmed that torus size luminosity relation with $R \propto L^{0.42\pm0.05}$.
Finally from both fitting the relations for W1 and W2 radii and using individual sources the temperature profile of the dust torus is estimated. From individual measurements this ranges in $\alpha$ from 0.3--1.1 and 0.65 from our relations. The predominantly shallow temperature profiles found in the majority of our sources are consistent with different dust distribution models. This demonstrates the potential of variability/reverberation mapping as a new approach to measuring the temperature structure of the AGN dusty torus

\end{itemize}

\begin{acknowledgments}
N. K. and X. D.
would like to acknowledge NASA funds 80NSSC22K0488, 80NSSC23K0379 and NSF fund AAG2307802. We thank
Las Cumbres Observatory and its staff for their continued support of ASAS-SN. ASAS-SN is funded by Gordon and
Betty Moore Foundation grants GBMF5490 and GBMF10501 and the Alfred P. Sloan Foundation grant G-2021-14192. This publication also makes use of data products from NEOWISE, which is a project of the Jet Propulsion Laboratory/California Institute of Technology, funded by the Planetary Science Division of the National Aeronautics and Space Administration.
\end{acknowledgments}

\begin{contribution}
All authors contributed equally.

\end{contribution}
\facilities{ASAS-SN, NEOWISE, Swift}
\bibliography{sample701}{}
\bibliographystyle{aasjournalv7}



\end{document}